\documentclass[pdflatex,sn-mathphys-num]{sn-jnl}

\usepackage{graphicx}%
\usepackage{multirow}%
\usepackage{amsmath,amssymb,amsfonts}%
\usepackage{amsthm}%
\usepackage{mathrsfs}%
\usepackage[title]{appendix}%
\usepackage{xcolor}%
\usepackage{textcomp}%
\usepackage{manyfoot}%
\usepackage{booktabs}%
\usepackage{algorithm}%
\usepackage{algorithmicx}%
\usepackage{algpseudocode}%
\usepackage{listings}%

\theoremstyle{thmstyleone}%
\theoremstyle{thmstyletwo}%

\theoremstyle{thmstylethree}%

\begin{document}

\title[Purified Giese-Salt]{Fine and Hyperfine Interactions with Multi-level Spin Relaxation of the purified \textit{Giese-Salt} in Veterinary Medicine: Prussian Blue Compound Ammonium-Ferric-Hexacyano-Ferrate}


\author*[1]{\fnm{Sascha A.} \sur{Bräuninger}}\email{Sascha.Albert.Braeuninger@tiho-hannover.de}
\equalcont{These authors contributed equally to this work.}
\author[1,2]{\fnm{Damian A.} \sur{Motz}}\email{motz@isah.uni-hannover.de}
\equalcont{These authors contributed equally to this work.}

\author[3]{\fnm{Sebastian} \sur{Praetz}}\email{sebastian.praetz@tu-berlin.de}
\author[4]{\fnm{Felix} \sur{Seewald}}\email{felix.seewald@tu-dresden.de}
\author[5]{\fnm{Katharina} \sur{Strecker}}\email{Katharina.Strecker@chemie.tu-freiberg.de}
\author[5]{\fnm{Carla} \sur{Vogt}}\email{Carla.Vogt@chemie.tu-freiberg.de}
\author[4]{\fnm{Hans-Henning} \sur{Klauss}}\email{henning.klauss@tu-dresden.de}
\author[3]{\fnm{Birgit} \sur{Kanngie\ss er}}\email{birgit.kanngiesser@tu-berlin.de}
\author[1]{\fnm{Hermann} \sur{Seifert}}\email{Hermann.Seifert@tiho-hannover.de}

\affil*[1]{\orgdiv{Institute of General Radiology and Medical Physics}, \orgname{University of Veterinary Medicine Hannover Foundation}, \orgaddress{\street{Bischofsholer Damm 15}, \city{Hannover}, \postcode{30173}, \state{Lower Saxony}, \country{Germany}}}

\affil[2]{\orgdiv{Institute of Sanitary Engineering and Waste Management}, \orgname{Leibniz University Hannover}, \orgaddress{\street{Welfengarten 1}, \city{Hannover}, \postcode{30167}, \state{Lower Saxony}, \country{Germany}}}

\affil[3]{\orgdiv{Institute of Physics and Astronomy}, \orgname{Technische Universität Berlin}, \orgaddress{\street{Hardenbergstraße 36, Sekr. EW 3-1}, \city{Berlin}, \postcode{10623}, \state{Berlin}, \country{Germany}}}

\affil[4]{\orgdiv{Institute of Solid States and Materials Physics}, \orgname{Technische Universität Dresden}, \orgaddress{\street{Haeckelstraße 3}, \city{Dresden}, \postcode{01069}, \state{Saxony}, \country{Germany}}}

\affil[5]{\orgdiv{Institute of Analytical Chemistry}, \orgname{TU Bergakademie Freiberg}, \orgaddress{\street{Lessingstraße 45}, \city{Freiberg}, \postcode{09599}, \state{Saxony}, \country{Germany}}}


\abstract{Ammonium ferric hexacyanoferrate is a veterinary-medical milestone and antidote against radiocesium, well-known as \textit{Giese-salt} after the Chernobyl disaster fed to domestic and wild animals, which shows even a rich interplay of properties in nanostructural chemistry and ferromagnetism. Among the broad analytical techniques, the ambivalence of macroscopic micrometer-sized agglomerates and nanoparticle sizes, a suggested enlarged Fe(II)$-$C$\equiv$N$-$Fe(III) bond length by Fe K-edge XAFS results and multi-level spin relaxation in $^{57}$Fe M\"ossbauer spectroscopy are highlighted. This sets this underestimated compound in a new light, e.g., for modern biomedicine and biofunctionality, extending its essential importance in addition to hypothetical future nuclear incidents.}




\maketitle

 \section*{Introduction}\label{intro}
The first synthesis of classic Prussian blue (PB, Berlin blue) by the paint manufacturer Johann Jacob von Diesbach took place between 1704 and 1706 \cite{Kraft2022, Gade1998_Kap2, Beyer2012Kap1}. This is often referred to as one of the most important discovery landmarks in coordination chemistry \cite{Buser1972, Buser1977, Ludi1981, Gade1998_Kap2, Beyer2012Kap1}. Among other aspects, the reasons for this pronounced significance are the unique physicochemical characteristics of this compound. One of these is the high thermodynamic and kinetic stability of PB as well as, closely related to this, the small solubility product \textit{K}$_\mathrm{sp}$(Fe$_4$[Fe(CN)$_6$]$_3$) $\approx$ 3 $\cdot$ 10$^{-41}$ \cite{Lurie1978, Nobrega1996}\footnote{The presented \textit{K}$_\mathrm{sp}$ value refers to the thermodynamic activities \textit{a} of the ions (see Lurie 1978 \cite{Lurie1978}). Thus, it is a dimensionless quantity.} and, thus, distinct low solubility in a relatively wide pH range \cite{Buser1972, Samain2013, Kjeldgaard2021} which makes it stable under (moderately) acidic, neutral and weakly alkaline conditions. The decomposition of PB occurs under alkaline conditions at pH $\geq$ 8 \cite{Latscha2004, Adhikamsetty2009, Manabe2020} and especially at pH $\geq$ 11 \cite{GIESE1988363}, into, for example, iron(III) hydroxide or oxide-hydroxide and hexacyanoferrate(II), as well as under hot and strong acidic conditions \cite{Roth2022, Welgama2023} that release hydrogen cyanide (HCN, prussic acid).
 
 Furthermore, PB can be prepared in two different variants, the "soluble" and the "insoluble" PB \cite{Kraft2022, Buser1977, Gervais2013, Samain2013, Wiberg+2008+1635+1680, Welgama2023}. With respect to the common classic laboratory and pharmaceutical scale direct synthesis routes of the compound by conversion of iron(III) cations and hexacyanoferrate(II) anions (ferrocyanide, as potassium salt\footnote{The sodium / ammonium salts are also feasible. In case of the "soluble" variant, the Na$^+$ / NH$_4^+$ derivative is obtained. In some literature, the Na$^+$, K$^+$ and NH$_4^+$ product versions are summarized just as Prussian blue and are not differentiated (e.g., \cite{Kraft2022, Estelrich2021}). Others differentiate the classic K$^+$-PB and the Na$^+$- / NH$_4^+$-PB and summarize them as Prussian blue pigments or even number the latter ones among Prussian blue compounds or analogues (e.g., \cite{Manabe2020}).}) or iron(II) cations and hexacyanoferrate(III) anions (ferricyanide, as the potassium salt) in aqueous solution \cite{Kraft2022, Estelrich2021, Samain2013,Wiberg+2008+1635+1680}, "soluble" PB is obtained in case of an equimolar educt ratio (and a preferably high dilution grade) or an excess of the hexacyanoferrate is used \cite{reguera_mossbauer_1992, Samain2013, Curdt2022, Riedel2007, Wiberg+2008+1635+1680, Welgama2023}. In fact, this so-called "soluble" version is a colloidal PB (sub-micron particle / crystal sizes), which can be described by the idealized chemical formula
 \begin{equation*}
 \mathrm{K}\{\mathrm{Fe(III)[Fe(II)(CN)}_6]\}
 \end{equation*}
 which is sometimes also referred to as K\{Fe(III)[Fe(II)(CN)$_6$]\} $\cdot$ x H$_2$O with x $=$ 1 - 5 \cite{Ware2008, Samain2013, Riedel2007} (see also Table \ref{tab0} and \cite{Kraft2022, Buser1977, Samain2013, Curdt2022, Ware2008, Wiberg+2008+1635+1680, Riedel2007, Welgama2023}). In contrast, direct synthesis routes yield "insoluble" PB, if an (large) excess of iron(III) or iron(II) cations is applied in the reaction \cite{reguera_mossbauer_1992, Samain2013, Curdt2022, Wiberg+2008+1635+1680, Riedel2007, Welgama2023, Carniato2020}. In this case PB precipitates in the form of a filtrable aggregated / agglomerated nano to micro crystalline solid, which is often referred to by the stylized chemical formula version based on the studies of Ludi \textit{et al.}
 \begin{equation*}
 \mathrm{Fe(III)}\{\mathrm{Fe(III)[Fe(II)(CN)]_6}\}_3 \cdot \mathrm{x}\;\mathrm{H_2O}
 \end{equation*}
(respectively Fe(III)$_4$[Fe(II)(CN)$_6$]$_3$ $\cdot$ x H$_2$O, see Table \ref{tab0}) with x $=$ 14 - 16 \cite{Kraft2022, Buser1972, Buser1977, Wiberg+2008+1635+1680}. Thus, PB is systematically called (potassium) iron(III) hexacyanoferrate(II), iron(III) ferrocyanide, or ferric ferrocyanide \cite{Kraft2022, Buser1972, Buser1977}. It should be mentioned that both variants of PB are also accessible by applying indirect synthesis routes (two-step procedures, where iron(II) cations with hexacyanoferrrate(II) form Prussian / Berlin white in a first step which is oxidized to PB in a second step), which are mostly used for large industrial-scale PB syntheses such as pigments \cite{Kraft2022, Samain2013}. 

\begin{table*}
\caption{Overview of common accepted chemical formulas, descriptions and abbreviations of some Prussian Blue Compounds (PBCs) as used in the literature and in this study. The mentioned literature references are just examples (more references are presented in the text). For the still on-going discourse about the correct / universal formulas, see text. A: alkali metal / ammonium cation, M, M*: 3d transition metal cations.}\label{tab0}
\begin{tabular}{lcc}
\toprule
Chemical Formula& Description/Synonyms/Abbreviation \\
\midrule
A$_x$M$_y$[M*(CN)$_6$]$_z$ $\cdot$ n H$_2$O \cite{Estelrich2021}&general formulas of entire class of\\
M$_x$[M*(CN)$_6$]$_y$ $\cdot$ n H$_2$O \cite{doi:10.1021/ic50206a032, Sharma2014}& Prussian Blue Compounds/Analogues,\\
&\textbf{PBCs/PBAs}\\
\cr
Fe(III)$\{$Fe(III)[Fe(II)(CN)$_6$]$\}_3$ $\cdot$ x H$_2$O&formulas of classic\\
Fe(III)$_4$[Fe(II)(CN)$_6$]$_3$ $\cdot$ x H$_2$O \cite{Kraft2022, Buser1972, Buser1977, Wiberg+2008+1635+1680}& "\underline{insoluble}" Prussian Blue, \\
Fe[Fe(CN)$_6$]$_{0.75}$$\square_{0.25}$ $\cdot$ x H$_2$O \cite{Kraft2022} &"\underline{insoluble}" \textbf{PB}\\
\cr
K\{Fe(III)[Fe(II)(CN)$_6$]\} ($\cdot$ x H$_2$O) \cite{Kraft2022, Samain2013, Wiberg+2008+1635+1680} & classic "\underline{soluble}" Prussian Blue (K$^+$),\\
&"\underline{soluble}" \textbf{PB}\\
\cr
NH$_4$\{Fe(III)[Fe(II)(CN)$_6$]\} ($\cdot$ x H$_2$O) \cite{GIESE1988363, Kaikkonen2000} & "\underline{soluble}" Prussian Blue (NH$_4$$^+$),\\
&\textbf{AFCF}, \\
& in veterinary medical applications\\ 
&also known as \textit{Giese-salt}\\
\bottomrule	
\end{tabular}
\end{table*}
 
\begin{figure}%
\centering
\includegraphics[width=\textwidth]{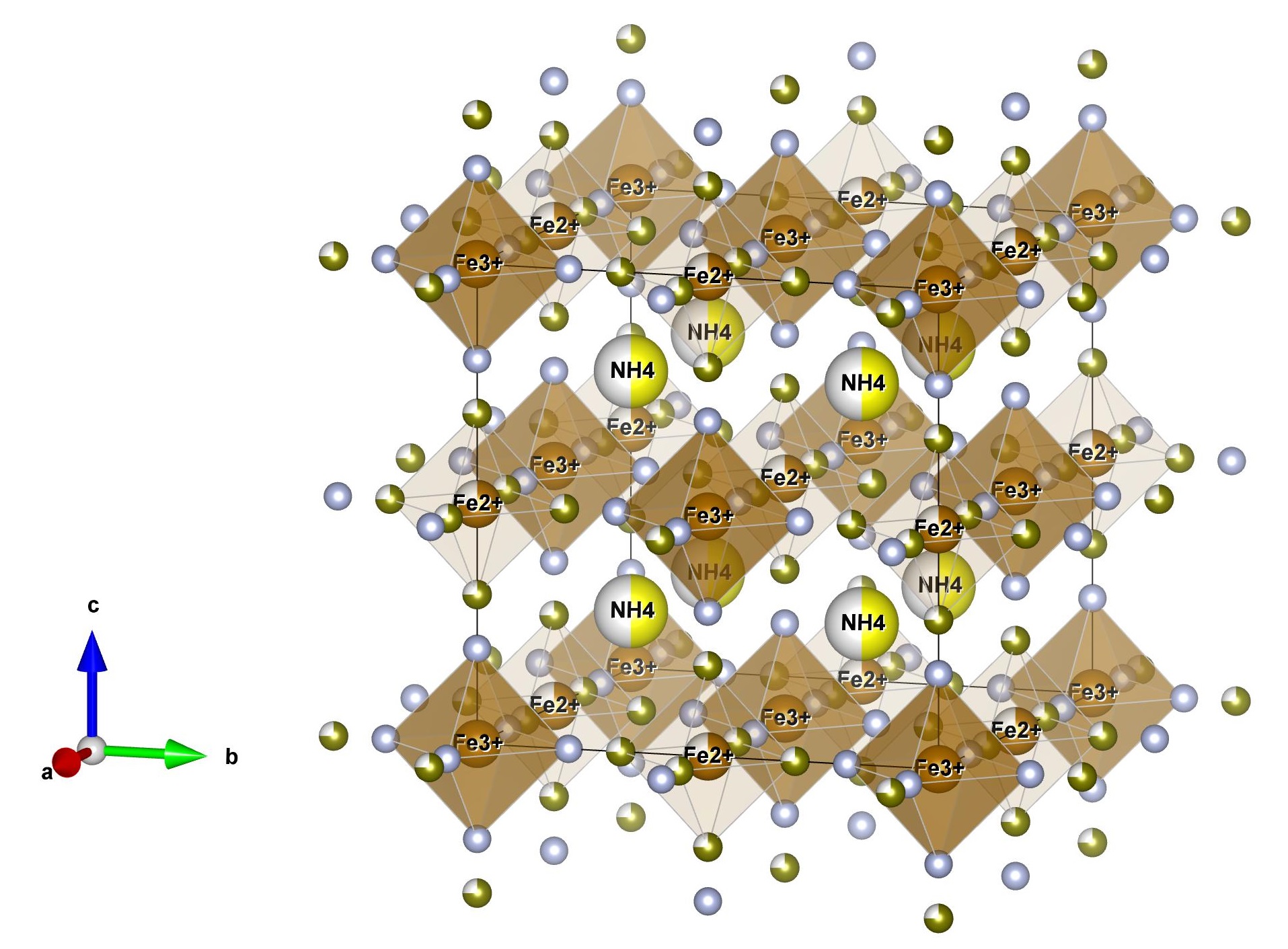}
\caption{Hypothetic simplified cubic crystal structure of Prussian blue Fe[Fe(CN)$_6$]$_{0.75}\square_{0.25}$ and corresponding enhanced unit cell (grey lines), H$_2$O molecules are not shown. C is shown in green, Fe is brown, N grey and NH$_4^+$ (or alkali metal cations) are yellow. 1/4 of the [Fe(CN)$_6$]$^{4-}$ are missing as emphasized by the missing probability (fractionalized white/brown balls of Fe$^{2+}$, fractionalized green balls, and, just half of the ammonium sites are occupied in case of ammonium-ferric-hexacyano-ferrate. The white probability means a probability for an empty position. Polyhedra emphasize the octahedral coordination of Fe as discussed. The structure is orientated on F. Herren \textit{et al.} 1980 \cite{doi:10.1021/ic50206a032}. \textit{VESTA} was used for visualization \cite{Momma:db5098}.}\label{figGie}
\end{figure}

 Both forms of PB feature a very intensive dark blue color, which is another example of the specific characteristics of this compound and is, in combination with a very low toxicity, the reason why PB has been widely used as a pigment (e.g., artists' color, laundry blue and also in analytics and bio-analytics as a product of color / staining reactions for sensitive qualitative Fe(III) or Fe(II) detection) since its discovery \cite{Kraft2022, Buser1977, Samain2013, Gervais2013, Wiberg+2008+1635+1680, Jander2006, SONODA2025}. These light absorption characteristics are traced back to an intervalence charge transfer between Fe(II) and Fe(III) in the mixed-valence compound \cite{Estelrich2021, Samain2013, Gervais2013, Wiberg+2008+1635+1680} and therefore to the solid-state structure of PB. The basic structure of both "soluble" and "insoluble" PB can be described with a cubic unit cell (Fig. \ref{figGie}) consisting of eight octants \cite{Kraft2022, Buser1972, Buser1977, doi:10.1021/ic50206a032, Sharma2014, Estelrich2021, Janiak+2007+381+580, Wiberg+2008+1635+1680, Riedel2007} and as a three-dimensional coordination polymer / network, which is similar to metal-organic frameworks (MOFs) \cite{Buser1977, Ludi1981, doi:10.1021/ic50206a032, Mamontova2022, Janiak+2007+381+580, Wiberg+2008+1635+1680}. In this so-called Keggin-Miles structure, both Fe(II) and Fe(III) are in octahedral coordination by cyanide (CN$^-$) anions \cite{Kraft2022, Buser1972, Buser1977, doi:10.1021/ic50206a032, Janiak+2007+381+580, Wiberg+2008+1635+1680, Riedel2007}. In detail, the carbon side binds Fe (II) (CN$^-$ as cyanide ligand) and the nitrogen side Fe (III) (CN$^-$ as isocyanide ligand) and, thus, chains of bridged Fe(II) and Fe(III) in the form of Fe(II)$-$C$\equiv$N$-$Fe(III) are formed \cite{Kraft2022, Buser1972, Buser1977, doi:10.1021/ic50206a032, Wiberg+2008+1635+1680, Riedel2007}. Since cyanide is a strong ligand and isocyanide is a weak ligand, this results in ground-state electronic configurations of Fe(II) in low-spin (LS) (3d$^6$: t$_{2\mathrm{g}}^6$e$_\mathrm{g}^0$ or $^1$A) and Fe(III) in high-spin (HS) state (3d$^5$: t$_{2\mathrm{g}}^3$e$_\mathrm{g}^2$ or $^6$A) \cite{Kraft2022, Buser1977, doi:10.1021/ic50206a032, Riedel2007}. 
 
Nevertheless, in further detail the structure and closely related to this, the compositional and structural differences between "soluble" and "insoluble" PBs are more complex and, in several aspects, are still part of recent research. Thus, one suggestion is that the "soluble" version is very close to the ideal structure mentioned above and contains a potassium cation due to charge compensation in every second octant \cite{Kraft2022, Riedel2007, Welgama2023}. In contrast, the "insoluble" form exhibits a structure containing statistically distributed [Fe(CN)$_6$]$^{4-}$ defects (1/4 of the [Fe(CN)$_6$]$^{4-}$ are missing) as well as a large number of water molecules. These water molecules are of three different types: coordinated H$_2$O  (3/4 of the Fe(III) are coordinated by four CN$^-$ and two H$_2$O), non-coordinated H$_2$O in the cavities / [Fe(CN)$_6$]$^{4-}$-vacancies (connected by hydrogen bonds to the coordinated H$_2$O) and additional non-coordinated water in the octants \cite{Kraft2022, Buser1972, Buser1977, doi:10.1021/ic50206a032, Sharma2014, Estelrich2021, Wiberg+2008+1635+1680, Riedel2007, Welgama2023}. Therefore, when intending also to consider and reflect the vacancies and water, the formula for "insoluble" PB should be written as Fe[Fe(CN)$_6$]$_{0.75}$$\square_{0.25}$ $\cdot$ x H$_2$O \cite{Kraft2022} (see Fig. \ref{figGie} and Table \ref{tab0}). 

Contrary, other studies and theories describe that "soluble" PB could also contain defects (in this case Fe(II) and Fe(III) defects in an equal ratio) or that both "soluble" and "insoluble" PB are basically of the same stoichiometry (K$\{$Fe(III)[Fe(II)(CN)$_6$]$\}$) and that the differences stem from just particle / crystal sizes (nanometer scale in the case of "soluble" PB) and additional [Fe(CN)$_6$]$^{4-}$ anions at the particle surfaces \cite{Kraft2022}. Here, the classic Ludi "insoluble" PB formula corresponds to a special case in the context of a very defect- and water-enriched "insoluble" form obtained under extreme reaction conditions and the absence of potassium \cite{Kraft2022}. However, there are also studies, e.g. Bueno \textit{et al.} 2008 \cite{Bueno2008} and Samain \textit{et al.} 2013 \cite{Samain2013}, that report "soluble" and "insoluble" PB are basically both of the "insoluble" stoichiometry (Fe(III)$_4$[Fe(II)(CN)$_6$]$_3$ $\cdot$ x H$_2$O), thus, both showing vacancies, but the "soluble" form just contains some inserted potassium cations (in total a stoichiometry of the type K$_\mathrm{y}$$\{$Fe(III)$_4$[Fe(II)(CN)$_6$]$_3$\} $\cdot$ $\{$y OH + x H$_2$O$\}$). 

The reasons for this persistent vagueness about the PB structure are diverse. A typical technique for detailed structural analyses of inorganic compounds is X-ray crystallography / X-ray diffraction (XRD) in the form of single-crystal analyses. Due to the very low solubility and generally (under standard conditions of the classic direct synthesis routes) fast precipitation, it is challenging to obtain PB single crystals or crystals of sufficient size and quality for such examinations \cite{Buser1972, Buser1977, Dostal1996, Samain2013}. In case of the "insoluble" PB, single-crystal preparations as well as X-ray crystallographic examinations were famously reported, for example, in Buser $\&$ Ludi 1972 \cite{Buser1972} and Buser \textit{et al.} 1977 \cite{Buser1977}. However, reaction and crystal growth, based on direct or indirect synthesis routes, require extreme or at least special conditions and procedures \cite{Buser1972, Buser1977}. With regard to the "soluble" version, some successful preparations of well-crystallized products have also been reported. For instance, Samain \textit{et al.} 2013 \cite{Samain2013} obtained the ammonium derivative (see below) of the classic "soluble" PB in a well-crystallized form by reproducing the conditions Buser $\&$ Ludi 1972 \cite{Buser1972} and Buser \textit{et al.} 1977 \cite{Buser1977} described for "insoluble" PB single crystals (based on the indirect synthesis path), but using an equimolar educt ratio. In addition to XRD as well as neutron diffraction, which has also been used in PB examinations \cite{doi:10.1021/ic50206a032, Sharma2014}, $^{57}$Fe-Mössbauer spectroscopy is another feasible method of structural analysis of PB (though only focused on the iron sides and the local environment), which has proven its suitability with regard to this sample system multiple times. The most popular example of successful application in PB research was the verification that Prussian blue and Turnbull´s blue correspond to each other and are basically identical \cite{reguera_mossbauer_1992, Ware2008, Samain2013, Riedel2007}. In general Mössbauer spectroscopy is prevalent in research regarding structural examinations of PB as well as the iron-containing examples of the derivational Prussian blue analogues / compounds (PBAs / PBCs: A$_\mathrm{x}$M$_\mathrm{y}$[M*(CN)$_6$]$_\mathrm{z}$ $\cdot$ n H$_2$O \cite{Estelrich2021} or M$_\mathrm{x}$[M*(CN)$_6$]$_\mathrm{y}$ $\cdot$ n H$_2$O \cite{doi:10.1021/ic50206a032, Sharma2014}, A: alkali metal / ammonium cation, M, M*: 3d transition metal cations; see also Table \ref{tab0}). However, the numerous Mössbauer studies also revealed a relatively large range and variability of the hyperfine parameters of PBCs. Although hyperfine parameters are generally known to be very sensitive to several sample properties (e.g. purity, particle sizes, crystallinity) and always show some sample variability \cite{Gütlich2013, Vandenberghe2013}, the parameters of PBCs exhibit especially large variations and can often even distinctively depend on the details of chemical synthesis from different provenances \cite{Samain2013, reguera_mossbauer_1992}. It must be emphasized that similar pronounced variability is also observed in other characteristics of PBC such as color (thus UV-Vis interaction), X-ray absorption fine structure, particle sizes / morphologies and detailed elemental composition (for instance variability in water contents) as reported, e.g., in detail for potassium, sodium and ammonium variants of "soluble" PB by Samain \textit{et al.} 2013 \cite{Samain2013}.

Another essential characteristic of PB and PBCs is the ion exchange qualities, which is similar to the characteristics of MOF or zeolites \cite{Kraft2022, Samain2013, Estelrich2021, Mamontova2022}. Thus, the potassium cations (K$^+$) of the classical "soluble" and "insoluble" PB can be exchanged by cations of cesium (Cs$^+$) or thallium (Tl$^+$) \cite{Kraft2022, Janiak+2007+381+580}. However, especially in the case of "insoluble" PB the Cs and Tl binding capacity is even increased, because in addition to the described ion exchange mechanism (or in the case of an ideal "insoluble" PB ion incorporation) other aspects, such as ion exchange with hydronium ions (simplified H$^+$, from H$_2$O in the PB structure), adsorption and ion trapping, also contribute to the binding process \cite{MARTINEZALONSO2024, YANG2008, Altagracia-Martínez2012}. 
In total, a very selective fixation of cesium is feasible \cite{GIESE1988363,gieserev}. Furthermore, in this cesium exchange, a decreasing colloidal solubility is also observed in the case of PB variants that are "soluble" \cite{GIESE1988363}. A very low resorbability and toxicity under physiological conditions is the reason for the common application of "insoluble" PB as an orally administered medication for thallium or especially radiocesium (Cs-137, Cs-134) intoxication (particularly oral incorporation) in human medicine ("insoluble" PB containing medications: \textit{Radiogardase-Cs} and \textit{Antidotum Thallii Heyl} by \textit{Heyl Chemisch-pharmazeutische Fabrik GmbH \& Co. KG}) \cite{Kraft2022, Janiak+2007+381+580, MARTINEZALONSO2024, YANG2008, Altagracia-Martínez2012}.

In this study, the focus was on the PBC ammonium iron(III) hexacyanoferrate(II) (or ammonium ferric hexacyanoferrate(II), AFCF), which is also known as \textit{Giese-salt} \cite{GIESE1988363, Kaikkonen2000} and can be described by the idealized chemical formula
\begin{equation*}
 \mathrm{NH_4}\{\mathrm{Fe(III)[Fe(II)(CN)]_6}\}
 \end{equation*}
(see also Table \ref{tab0}). It is the ammonium derivative of the classic "soluble" PB. AFCF has been applied in veterinary medicine as a food additive for the decorporation of Cs-137 and Cs-134 from wild, domestic and farm animals and thus also to reduce the radiocesium burdens in animal-derived foods (e.g., meat and milk) for several decades and is still used today \cite{GIESE1988363,m_giese_1988,safetygiesesatl,DABURON199173, Kaikkonen2000}. The application related to the Chernobyl nuclear reactor accident in 1986, which led to major radiocesium fallout across Europe, remains one of the most notable and widespread. Especially due to the proportionally long radioactive half-life of $^{137}_{\:\,55}$Cs ($\tau$$_{1/2}$ $=$ 30.1\,a) \cite{Bethge2008} the application of AFCF (for example, for wild boars) is still partially requested. The \textit{Giese-salt} was also optimized for the use as additive in lick stones and, in the form of substrate-fixed AFCF, for the decontamination of whey. Historically, there was also a pigment of an ammonium PB derivative known as Monthier's blue and was first synthesized in 1846 \cite{Monthiers, riffault_des_hetres_practical_1874, Gournis2010, pigment} by reaction of an indirect synthesis route \cite{Monthiers, riffault_des_hetres_practical_1874, pigment, Robine1906}. However, an alternative synthesis path, based on the mentioned classic direct synthesis routes, in radiochemistry with the described purpose of product application in veterinary medicine naturalized the synonym \textit{Giese-salt} (german: \textit{Giese-Salz}\footnote{Nowadays, the name \textit{Giese-Salz} is used in the German and Scandinavian literature, main stream articles and radiation protection community labeled by W.W. Giese (*1934$-\dagger$2023).}). 

The purpose of this study was a detailed and expanded examination of AFCF intended for application in veterinary medicine by appropriate modern elemental and species analysis techniques. Hence, the motivation was to contribute to the goal of achieving an improved understanding of the chemical and structural characteristics of (veterinary) medical AFCF. Since PBCs show extraordinary variability in the structural details and characteristics as it has always been revealed in extensive studies (see above), the requirements for human and veterinary medical purposes in the sense of purity and corresponding chemical reproducibility, as expected for biomolecular applications, are stressed. An improved understanding is beneficial and required for further optimized examinations and elucidations of the pharmaceutical and biological effectiveness and mechanisms of AFCF. 

Industrial AFCF produced by the direct synthesis route and delivered for medical applications in veterinary medicine invariably contains large amounts of the by-product ammonium chloride of $\omega$(NH$_4$Cl)$_{\mathrm{rel}}$ $\approx$ 35 - 40$\%$ \cite{GIESE1988363, safetygiesesatl}. In contrast, the synthesized AFCF sample examined in this study was purified by dialysis to reduce or remove the by-product content \cite{GIESE1988363} and related analytical interferences and matrix effects. After several first macroscopic, chemical, microscopic and laser diffraction (LD) examinations, the sample was analyzed by the use of inductively coupled plasma optical emission spectroscopy (ICP-OES) and CHNS analysis to determine the elemental composition in detail. For in-depth species and structural examination, X-ray powder diffraction (XRD, or more precisely PXRD), vibrational spectroscopy (Raman and IR spectroscopy), UV-Vis/NIR spectroscopy, X-ray absorption fine structure spectroscopy (XAFS) and room-temperature as well as low-temperature $^{57}$Fe Mössbauer spectroscopy were applied. Furthermore, suitable reference substances were required especially for the XAFS measurement. Thus, in addition to commercially received potassium hexacyanoferrate(II) and hexacyanoferrate(III), three "classic" Prussian blues (two "insoluble" PBs and one "soluble" K$^+$-containing PB) were also synthesized and pre-characterized in the scope of this study.

\section*{Experimental}\label{experimental}

\subsection*{Sample system - syntheses and pre-characterizations}
\subsubsection*{Ammonium iron(III) hexacyanoferrate(II) (AFCF, \textit{Giese-salt})}
The \textit{Giese-salt} (in the following just referred to as AFCF) presented in this work had already been synthesized previously for this study and continuously stored in a glass vessel sealed with \textit{Parafilm}\textsuperscript{\textregistered}. A storage place avoiding direct sunlight exposure was also chosen. The synthesis was performed in the same way as described by Giese 1988 \cite{GIESE1988363}. Thus, based on the direct laboratory-scale synthesis routes for PBCs, AFCF was obtained by an equimolar reaction of ammonium hexacyanoferrate(II) ((NH$_4$)$_4$[Fe(CN)$_6$]) and iron(III) chloride (FeCl$_3$) in aqueous solution (see Scheme S1 in Supplementary Information SI). It should be noted that the synthesis also contained a product purification by means of a dialysis to reduce the by-product (ammonium chloride, NH$_4$Cl) and residual educt contents that provided the purity of the AFCF required for the examinations carried out in this study. 

\subsubsection*{Prussian blue reference substances}
Especially for XAFS examinations, suitable PB reference substances were required in addition to the commercially purchased reference potassium hexacyanoferrate(II) and potassium hexacyanoferrate(III) (see Experimental Subsection XAFS below). Thus, three PB reference substances were synthesized by different preparation variants based on the classic direct synthesis route (for further details on the syntheses see Supplementary Information SI). Because all three synthesized PBs were intended for application as reference substances, the synthesis variants were specifically developed or modified with a focus on purity of the obtained products. Therefore, sufficient cleaning procedures of the filtered and separated solid products were of interest (see SI). 

The obtained products were labeled as PB I, PB II, and PB III for the purpose of distinctness in the course of this study: 
\begin{enumerate}
    \item PB I: classic "insoluble" Prussian blue, Fe(III)$\{$Fe(III)[Fe(II)(CN)$_6$]$\}_3$ $\cdot$ x H$_2$O, synthesized by a two-step procedure via an \textit{in situ} formed "soluble" PB which is transferred into the "insoluble" variant. 
    \item PB II: classic "insoluble" Prussian blue, Fe(III)$\{$Fe(III)[Fe(II)(CN)$_6$]$\}_3$ $\cdot$ x H$_2$O, synthesized by the addition of a [Fe(CN)$_6$]$^{4-}$-solution to an Fe(III) solution of very high concentration (high Fe(III) excess).   
    \item PB III: classic "soluble" Prussian blue, K\{Fe(III)[Fe(II)(CN)$_6$]\} ($\cdot$ x H$_2$O), synthesized by an equimolar reaction of Fe(III) and [Fe(CN)$_6$]$^{4-}$ at high dilution grade.  
\end{enumerate}
Beforehand and between measurements and applications, the synthesized compounds PB I, II and III were stored in snap-cap vials sealed with \textit{Parafilm}\textsuperscript{\textregistered} and kept away from direct sunlight exposure. 

 PB I, II, and III were pre-characterized with respect to the required product identity and purity before being used as reference substances in the AFCF examinations. Thus, a colloidal dispersion test and ATR-IR spectroscopy were applied. The dispersion tests confirmed that both PB I and PB II showed typical "insoluble" characteristics (without and also with ultrasonic treatment a colloidal "solution" was not obtained)  whereas PB III exhibited "soluble" properties (see SI Fig. S1). The measured IR spectra (see SI Fig. S2 and Table S4) were in good accordance with the literature IR spectra of the PBCs on band positions, shapes, and intensities. Hence, the pre-characterizations confirmed the required identities and sufficient purities of the synthesized reference substances (for further details on the pre-characterization results, see SI).       
\subsection*{Rapid tests for chemical characteristics}
\subsubsection*{Colloidal dispersion test}
Colloidal dispersion tests were applied to differentiate between “soluble” and “insoluble” PBC variants. Therefore, these tests enabled a simple and fast first classification of the AFCF sample as well as the synthesized reference PBs (pre-characterizations of the synthesized reference PBs, see above) based on the colloidal dispersion characteristics. 

In a 50 ml beaker, a small amount (spatula tip) of the sample was mixed with 10 ml of demineralized water. The mixture was slewed slightly. A positive test result, which means the outcome of a (clear) blue “solution”, indicated the presence of a 'soluble' PBC, while a negative result proved the identity of an “insoluble” PBC. 

In some cases of “soluble” PB samples, an additional short (\textit{t} = 1 – 3\,min) treatment by an ultrasonic bath to obtain a virtually complete clear colloidal “solution” (due to particle agglomerations, especially in cases of samples that had been stored for longer periods of time before examinations) was required. Nevertheless, this procedure did not limit the expressiveness of the test, since “soluble” and “insoluble” PBC variants can still be distinguished. If an “insoluble” PB is present, even this short ultrasound bath treatment will not produce a complete colloidal dispersion (if any dispersion occurs at all). This was found for samples examined within the scope of this study, but has also been reported in the literature, e.g. by Samain \textit{et al.} 2013 \cite{Samain2013}.

\subsubsection*{Ammonium test}
In classic wet chemical qualitative analysis, samples can be tested for the presence of ammonium (NH$_4$$^+$) by the formation of (gaseous) ammonia (NH$_3$) under strong alkaline conditions due to a Br$\o$nsted-Lowry acid-base reaction \cite{Jander2006}. Since AFCF is an ammonium-containing PBC, this test was applied to the AFCF sample examined in this study. A small sample amount was mixed with solid sodium hydroxide (NaOH) and a few drops of demineralized water on a watch glass. A piece of moistened universal indicator paper was fixed on a second watch glass, which was then placed as a cover on the first. It must be emphasized that a direct contact between the indicator paper and the sample-NaOH-mixture must be prevented, since the goal is to test for gaseous ammonia, which is formed and released, if ammonium is present. If the indicator color turns blue after a few seconds, the presence of ammonium in the sample is proved.

\subsection*{Particle morphology and size examinations}

\subsubsection*{Light microscopy}
In this study, two types of light microscopy techniques were considered. 

Firstly, transmitted-light microscopic examinations were performed with an \textit{Axiostar Plus} (\textit{Carl Zeiss AG}) equipped with a digital camera \textit{Axiocam 105 color} (\textit{Carl Zeiss AG}). Objective lenses (\textit{Carl Zeiss AG}) with 4x and 10x magnification were used. For examinations, the AFCF sample was prepared using an adhesive tape technique. On a glass microscope slide, a piece of tape (\textit{tesafilm}\textsuperscript{\textregistered} \textit{kristall-klar}) was fixed with the adhesive side facing upwards. Very small amounts of the sample (a tipful of a micro spatula of substance) were evenly applied to the adhesive surface. Finally, the microscope slide was carefully tapped sideways on the surface of the table, to remove excess material and to get a preferably thin sample layer. For image capture as well as a first evaluation of the pictures the software \textit{ZEN lite} (version 3.3, \textit{Carl Zeiss AG}) was applied. Further processing was performed with \textit{ImageJ} (version 1.54g). 

Secondly, reflected-light microscopy was also applied since this technique is, for example, well suited for examinations of nontransparent samples, hence opaque samples \cite{Rochow1978, Leng2013}, but in particular also for powder samples \cite{Leng2013} such as the AFCF sample of this study. This is especially true when one also intends to analyze the particle shapes, color, and surface structure or topography. For examinations a light microscope of the type \textit{Motic SMZ-171 Trinokular} with 2x attachment lens, the camera \textit{Bresser MicroCam SP 5.0} and the software \textit{Bresser MicroCamLabII} was utilized. For sample preparation, a small amount of the AFCF sample was placed directly on a microscope slide. Analogously to transmitted-light microscopy, further evaluation and processing of the images was performed using the software \textit{ImageJ} (version 1.54g). 

\subsubsection*{Scanning electron microscopy}
Scanning electron microscopy (SEM) was done by using a \textit{VEGA SB} from \textit{TESCAN}. The measurements were performed using an accelerating voltage of \textit{U} = 20\,kV and the secondary electron detection (SED) mode. The AFCF sample was prepared on adhesive graphite pads that were attached to typical aluminum sample holders. As was also the case with the light microscopy images, the further evaluation and processing of the SEM images was conducted with \textit{ImageJ} (version 1.54g). 

\subsubsection*{Laser diffraction}
For a detailed particle size analysis of the AFCF sample laser diffraction (LD) was applied. These examinations were conducted with the \textit{Microtrac} particle analyzer \textit{SYNC}. This analyzer enables LD and dynamic image analysis (DIA) simultaneously. However, in this study, the sole application of LD was sufficient since the focus of these measurements was just on the determination of the volume-based particle size distribution. The goal was a comparison with LD-based particle size analysis results for an AFCF sample prepared and purified the same way as in this study which had been reported by Giese 1988 \cite{GIESE1988363}. The examinations were performed on a wet sample system, hence the AFCF in colloidal "solution" (after 60\,s of ultrasonic bath treatment). The measurements were carried out as duplicate determinations and the outcomes were averaged to get an overall result. The further evaluation was performed based on the specifications presented in \textit{DIN ISO 13320:2022-12} \cite{DIN-13320}, \textit{DIN ISO 9276-1:2004-09} \cite{DIN-9276-1} and \textit{DIN ISO 9276-2:2018-09} \cite{DIN-9276-2}.  

\subsection*{Elemental analysis}
\subsubsection*{Inductively coupled plasma optical emission spectroscopy}
Quantitative elemental analysis in the form of inductively coupled plasma optical emission spectroscopy (ICP-OES) was performed by using a \textit{Spectro ARCOS-SOP} (\textit{Spectro Analytical Instruments}). The analysis consisted of multiple steps and aspects. 

In a first step, the AFCF sample was dissolved by microwave assisted digestion at \textit{T} = 453.15\,K (\textit{t} = 10\,min) in diluted nitric acid (5\,mL pure H$_2$O + 1\,mL p. a. sub boiled 65\,wt.\% HNO$_3$). Three individual samples in the mass range of about 30 – 50~mg (sample 1: \textit{m} = 0.0486\,g, sample 2: \textit{m} = 0.0362\,g, sample 3: \textit{m} = 0.0227\,g) were dissolved (three replicate / triple determination). Additionally, a blank digestion was also performed considering possible contamination of the digestion chemicals used as well as the microwave tubes. In the second step, the obtained clear sample solutions and the blank solution were diluted with pure water to a total mass of about 40 g (sample 1: \textit{m} = 40.0220\,g, sample 2: \textit{m} = 39.7425\,g, sample 3: \textit{m} = 39.9339\,g, blank: \textit{m} = 40.2300\,g) for the further ICP-OES measurements. 

The elemental quantifications by ICP-OES were performed by external calibrations. Thus, multi-element commercial stock solutions were diluted, for matrix adjustment reasons, with 2.5\,wt.\% HNO$_3$ to obtain standard solutions. In the ICP-OES measurements of each examined element, three selected lines were considered. Line selections were based on several aspects such as sensitivity and interferences. The measurement results of each element line were evaluated based on \textit{DIN 38402-51:2017-05} \cite{DIN-38402}. These single line results were averaged. Finally, the element contents determined from the three AFCF samples were tested for outliers (Grubbs test) first and then also averaged to obtain overall results. 

Based on previous semi-quantitative measurements, in total thirteen elements were considered in the main ICP-OES evaluations (Ba, Ca, Co, Cr, Cu, Fe, K, Mg, Mn, Na, Ni, S and Zn), although the focus was primarily on the iron and potassium content because these two elements allowed crucial statements about the AFCF sample identity and purity, or more precisely specific composition (for example, regarding the water content). 

\subsubsection*{Elementary analysis (CHNS analysis)}
Additional quantitative elemental analysis with the focus on nonmetals was applied. This analysis was performed in the form of CHNS analysis, thus, the elements carbon (C), nitrogen (N), hydrogen (H), and sulfur (S) were considered in the examinations. For measurements a \textit{vario MICRO cube} (\textit{Elementar}) was used. In total, two individual samples of the AFCF sample were examined (double replicate determination), and the results obtained were averaged to obtain an overall result. 

\subsection*{X-ray powder diffraction}
XRD was performed in reflection geometry (Bragg-Brentano) by using a \textit{D8 Discover} X-ray diffractometer (\textit{Bruker}) with a Copper K$\alpha$ X-ray source ($\lambda$$_{Cu-K\alpha}$ = 1.5406\,\AA) and a compound silicon strip detector \textit{LYNXEYE} (\textit{Bruker}). The covered measurement range was between 5° and 70° 2$\theta$ with a step size of 0.020°. The AFCF powder sample was prepared flat on a sample holder. The measurement was performed at room temperature. For the qualitative phase analysis the software \textit{DIFFRAC.EVA} (\textit{Bruker}) and \textit{WinXPow} (version 1.08, \textit{STOE \& Cie GmbH}) were employed. The reason for the application of different software was simply that the evaluations of the measured raw data were performed at two research facilities simultaneously due to the availability and access options of different XRD data bases. Overall, no different results were expected. Indexing of the measured diffractogram was subsequently performed using the quadratic Bragg equation (see SI, equation (S1)) for cubic crystal systems \cite{spiess2009moderne, Borchardt-Ott2009}. 

For crystal size approximation the Scherrer equation based on the full-width-at-half-maximum-method (FWHM-method) \cite{spiess2009moderne, Sharma2012, Ameh2019} was applied (see SI, equation (S2)). In total, the first three most intensive reflections (hkl) = (200), (220), and (400) were considered in the calculations. These three individual results were averaged to obtain an overall result. However, it must be emphasized that this method had to be applied without an instrumental broadening correction because an experimental determination of this parameter could not be performed in the scope of this study. Since instrumental broadening can also notably contribute to the total broadening of reflections observed in an XRD, a correction of the observed FWHM is usually essential to obtain precise crystal size approximations \cite{spiess2009moderne, Ameh2019}. Hence, a more pronounced uncertainty / bias could be expected for the results obtained in this study.

\subsection*{Vibrational spectroscopy}

\subsubsection*{Raman spectroscopy}
Raman spectroscopy was performed using an \textit{XploRA Plus} (Raman Spectrometer: Confocal Raman Microscope) from \textit{Horiba Scientific} with a wavelength of a Nd:YAG laser of 532\,nm. For soluble and insoluble Prussian blue, a so-called ’safe zone’ for the laser power was defined by Moretti \textit{et al.} 2018 \cite{Moretti2018RamanSO} between 0.0005\footnote{Lower threshold is defined and not measured because of a high signal-to-noise-ratio.}-0.06\,mW. In this study, we used a laser power of 0.0126\,mW, below the critical threshold, avoiding a considerable frequency shift and an accumulation time of 60\,s as proposed \cite{Moretti2018RamanSO}. The data were calibrated using a silicon wafer. For the measurements the software \textit{LabSpec 6} from \textit{Horiba} was used. The Raman measurement data were smoothed and filtered by the Savitzky-Golay filter \cite{doi:10.1021/ac60214a047} to increase the precision of the data without distorting the trend of the signal.

\subsubsection*{Infrared spectroscopy}
The technique of attenuated total reflectance (ATR) was employed for infrared spectroscopy (IR) assessments at room temperature. These measurements were performed using the \textit{Nicolet iS5} instrument, manufactured by \textit{Thermo Scientific}, which was equipped with a \textit{iD7 ATR} cell also from \textit{Thermo Scientific}. The equipment was able to detect wave numbers ranging from 525\,cm$^{-1}$ to 4000\,cm$^{-1}$. For data collection and processing (e.g. background correction of the raw spectra) \textit{OMNIC Series Software} (\textit{Thermo Scientific}) was applied. 

\subsection*{UV-Vis/NIR spectroscopy}
UV-Vis spectra were collected at room temperature in aqueous solution / dispersion with a measuring range of $\lambda$ = 190\,nm – 1300\,nm and a step size of 0.5\,nm. An UV-Vis/NIR double-beam Spectrophotometer \textit{V-670} from \textit{Jasco} and 10\,mm QX quartz glass cuvettes (\textit{Hellma Analytics}) were used for the measurements. As measurement software \textit{Spectra Manager} (\textit{Jasco}) was utilized. For the examinations, dispersions of an AFCF sample concentration of about \textit{c}(AFCF) $\approx$ 5 $\cdot$ 10$^{-4}$\,mol/L were prepared. Since stable dispersions were required with preferably low particle light scattering (thus, the requirements for the grade of colloidal “solution” were slightly higher than for the colloidal dispersion test described above), different techniques were tested, e.g. pure water + slewing, pure water + ultrasonic + slewing, and diluted nitric acid + slewing. The best and fastest result for the AFCF sample investigated in this study was obtained using diluted nitric acid (\textit{c}(HNO$_3$) = 1\,mol/L). Therefore, these acidic very fine dispersions were used for the UV-Vis/NIR spectroscopy. Accordingly, pure diluted nitric acid (in a cuvette of the same type and characteristics as the one for the sample solution; see above) was applied for the reference beam path. 

\subsection*{X-ray absorption fine structure spectroscopy}
X-ray absorption fine structure (XAFS) measurements for the Fe K-edge were carried out with a self-developed laboratory-scale wavelength-dispersive spectrometer on von H\'{a}mos geometry \cite{Schlesiger2015, Schlesiger2020} which has already been frequently used in various qualitative and quantitative examinations of iron species \cite{Schlesiger2015, Motz2021, Motz2023, Gili2024, Praetz2025, Praetz2025_3, Praetz2025_2} and also investigations of other species of elements (for example, cerium species) \cite{Gili2024, Oliveira2023}. The spectrometer is equipped with a microfocus X-ray tube with molybdenum as anode material, a curved highly annealed pyrolytic graphite (HAPG) mosaic crystal, and a hybrid photon-counting CMOS detector with 512 x 1030 pixel and a pixel size of 75\,µm x 75\,µm. The tube was operated with a high voltage of 13.2\,kV and a current of 1730\,µA. 

The sample preparations of the AFCF as well as the required chosen reference substances (see below) were carried out as the pellet-technique. Therefore, samples were prepared by mixing sample powders with wax-based binder \textit{Hoechst Wax C} (\textit{Merck KGaA}) and grinding the mixture with a mortar to acquire sufficient sample homogeneity and particle size below 50\,µm. The mixture was then pressed as a pellet with 13\,mm diameter and fixed between adhesive tape. The prepared sample pellets were measured in transmission at room temperature. It must be highlighted that two (or in some cases even three) sample pellets of each sample were prepared and measured to verify reproducibility and also to preclude effects of sample inhomogeneity. Furthermore, each sample pellet was also measured two or, in cases of all PBCs, three times. During the measurements, the sample pellets were constantly moved to minimize the effects of local thickness inhomogeneity. The measurement time varied between 5\,h and 10\,h for each sample. For energy axis calibration purposes, an $\mathrm{\alpha}$-Fe reference (foil and/or powder) was measured before the sample examinations. For normalization of the data obtained and further analysis of the spectra \textit{ATHENA} of the Demeter software package has been used \cite{Ravel2005}. Although both the X-ray absorption near-edge structure (XANES) and the extended X-ray absorption fine structure (EXAFS) ranges of the obtained XAFS spectra were considered in the evaluations of this study, the focus was more on the XANES (including pre-edges and edges). Fe K-edge positions $E_0$ were determined by applying a bisect function on the normalized spectra to find the corresponding energy value to the value of $\mu$$Q$ = 0.5. 

In total, two commercially available and three self-synthesized materials (details see above and SI) were used as reference substances in the XAFS measurements: 
\begin{enumerate}
    \item Potassium hexacyanoferrate(II) trihydrate: K$_4$[Fe(II)(CN)$_6$] $\cdot$ 3 H$_2$O, crystalline, purity $\geq$ 98\%, \textit{Carl Roth}.
    \item Potassium hexacyanoferrate(III): K$_3$[Fe(III)(CN)$_6$], powder, purity 99\%, \textit{Sigma Aldrich}.
    \item PB I: synthesized in scope of this study, just as
    \item PB II, and
    \item PB III.
\end{enumerate}

\subsection*{$^{57}$Fe M\"ossbauer spectroscopy}
For the $^{57}$Fe M\"ossbauer spectroscopy, a \textit{WissEl} M\"ossbauer spectrometer (\textit{Wissenschaftliche Elektronik GmbH}) was used, with detection performed by a proportional counter tube or a Si-PIN detector from \textit{KETEK}. All measurements were performed in transmission with a Rh/Co source (the 14.4\,keV $^{57}$Fe-$\mathrm{\gamma}$-line was used for the measurements). The source started with an initial activity of 1.4\,GBq. $\mathrm{\alpha}$-Fe was used for calibration. The sample measurements were performed at room temperature as well as at lower temperatures (down to \textit{T} = 2.5\,K). The low-temperature $^{57}$Fe M\"ossbauer spectroscopic examinations were conducted using helium flow cryostats from \textit{Cryo Vac} and \textit{Oxford Instruments}, operated in under-pressure and normal mode. The helium reservoir was protected by a liquid nitrogen shield. \textit{Moessfit} software \cite{kamusella} was utilized for the analysis including quantum mechanical relaxation.

\section*{Results and discussion}\label{results}
\subsection*{Macroscopic description and specific chemical characteristics}
For a first verification of the identity of the synthesized purified AFCF and in the scope of this study, several macroscopic and chemical examinations were performed. The confirmation of the basic product identity was an essential requirement for all further examinations, especially detailed elemental and species analysis, which is the focus of this study.

\subsubsection*{General description}
The synthesized and purified AFCF sample was on hand as an intense dark-blue and fine powdery solid (see SI Fig. S3). The sample revealed a slight tendency towards electrostatic charging, which is characteristic of solids consisting of particles with very small sizes in the low-micrometer and especially the sub-micrometer ranges \cite{Deng2023}. Therefore, the basic characteristics regarding the morphology and color fitted to those of PBCs or more specific PB pigments.

\subsubsection*{Colloidal dispersion test} 
\begin{figure}
\centering
\includegraphics[width=\textwidth]{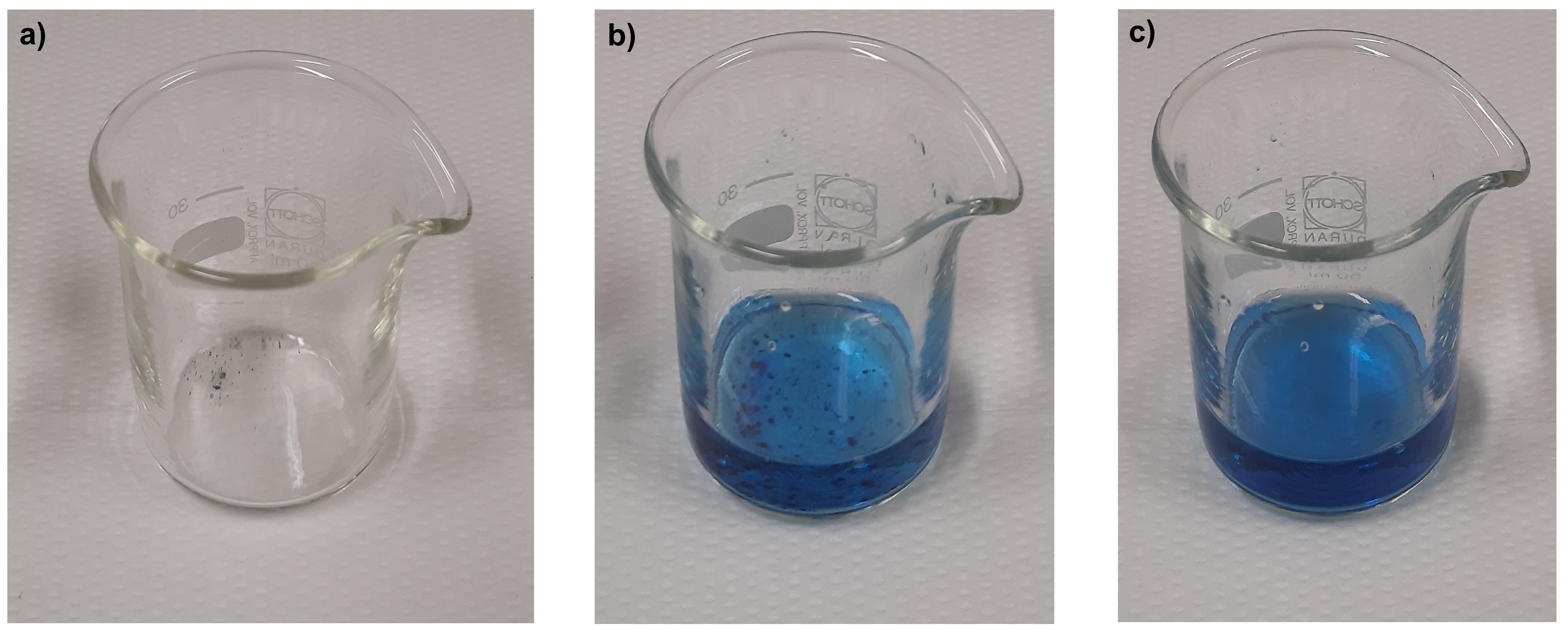}
\caption{Test for colloidal dispersion characteristics of the AFCF sample: a) Small amounts of the solid sample were placed in a beaker. b) 10 mL demineralized H$_2$O were added. c) Slight slewing quickly resulted in a clear blue colloidal dispersion. However, in some test runs a short ultrasonic treatment was required for a complete / clean "solution".}\label{fig:loesung}
\end{figure}

For a first species analysis of the AFCF sample, the colloidal dispersion test was performed. This test gave a positive result and a clear blue colloidal dispersion (see Fig. \ref{fig:loesung} and SI Fig. S4) was obtained, although in some test runs a short treatment by an ultrasonic bath (see Experimental) was required for a virtually complete "solution". This indicated the presence of agglomerates in the sample which was verified by the microscopic and LD examinations in the further analyses (see below). Altogether, the test confirmed the expected colloidal “solubility” of a “soluble” PBC and was a first validation of sample identity, since AFCF belongs to the "soluble" PBCs.

\subsubsection*{Qualitative wet-chemical ammonium test}
Furthermore, a fast qualitative test for the presence of ammonium in the AFCF sample was applied. The positive result (see SI Fig. S5) confirmed the presence of ammonium in the sample and therefore the presence of ammonium-containing PBC. 

In total, the three performed macroscopic and chemical tests confirmed the required basic product identity of the sample as an AFCF and justified the further detailed examinations and characterizations presented in the following sections.

\subsection*{Particle morphologies and sizes}
To verify the detailed particle size analysis by light scattering (laser diffraction analysis) which had already been reported by Giese 1988 \cite{GIESE1988363} for the AFCF sample synthesized and purified the same way as in this study (see SI) and to analyze the particle morphology, light microscopy, scanning electron microscopy (SEM) and laser diffraction (LD) were performed. Regarding the microscopic examinations, it should be noted that a precise determination of the number-based particle size quantity distribution was not performed and that the microscopic analyzes were only used for an overall check of the particle sizes. However, a precise volume-based particle size distribution analysis was performed with LD. Furthermore, an approximation of the crystal size of the XRD analysis of this study was also performed (see below). As can be clearly seen in light microscopy and SEM images in SI Fig. S6, the AFCF sample consists of agglomerates exhibiting a bulky morphology similar to the shapes widely reported in literature, e.g. by Samain \textit{et al.} 2013 \cite{Samain2013} for different synthesized and commercially available "soluble" and "insoluble" PB pigments. The particle sizes are mostly in the magnitude \textit{d} $\leq$ 100\,µm or, in detail, in range of \textit{d} $\approx$ 5 - 100\,µm, although some smaller particles in the submicrometer range (presumably non-agglomerated grains) but also larger particles (\textit{d} $\geq$ 100\,µm, see Fig. S6f and S6g) can be found as well. 

Similar results were also obtained for the detailed particle size distribution analysis based on LD. In the obtained volume-based particle size distribution density $q_3(d)$ and normalized distribution sum $Q_3(d)$ (see SI Fig. S7) it is clearly visible that the majority of the particles (agglomerates) in the AFCF sample are in the broad size range of \textit{d} $\approx$ 5 - 400\,µm. An arithmetic mean for the particle size of $\overline{d}$ $\approx$ 118\,µm $\pm$ 123\,µm (the uncertainty is the standard deviation) was determined from the measured distribution. However, as also evident by the high standard deviation, the obtained particle size distribution differed distinctively from a normal distribution, which limits the expressiveness of the arithmetic mean mentioned. This is also true for the log-normal distribution (which is in many cases an optimized description of particle size distributions \cite{DIN-9276-2}) and the related geometric mean, since the measured size distribution of the AFCF sample corresponded more to a bimodal-like distribution with maximums at $d_{max1}$ $\approx$ 41\,µm and $d_{max2}$ $\approx$ 227\,µm whereby the maxima hardly emerge from the overall distribution when lows and errors are taken into account (see SI Fig. S7). In total, these results are in good agreement with the findings of Giese 1988 \cite{GIESE1988363}, where a particle size distribution in a similar broad range of \textit{d} $=$ 5 - 500\,µm with an average and maximum at \textit{d} $\approx$ 30 - 60\,µm has been reported. The occurrence of a second maximum at larger particle sizes (or more precisely, agglomerates), which may be overinterpreted in the context of actual significance, could be caused by slightly different storage conditions, and the times beforehand the particle size examinations, do not limit the significance of molecular considerations.

\subsection*{Elemental analysis}
The primary goal of the elemental analysis was the qualitative and quantitative determination of all elements (except oxygen) expected to be contained in pure AFCF, hence carbon (C), iron (Fe), hydrogen (H), and nitrogen (N) and, therefore, to confirm or determine the chemical composition or formula, respectively. In addition, investigations of impurities (contained trace elements) of the AFCF sample were the secondary goal. The most important results obtained from the elemental analysis by ICP-OES and CHNS analysis are shown compared to the theoretical values in Table \ref{tab2} (for more results and details see SI Table S1).
\begin{table}
\caption{Exemplary results for elemental analysis of the AFCF sample (for further results, see SI Table S1) in comparison to ideal theoretical values (assuming the typical water contents x = 0, 1, 2, 3, 4 and 5 in the chemical formula NH$_4\{$Fe(III)[Fe(II)(CN)$_6]\}$ $\cdot$ x H$_2$O described in the literature for classic "soluble" PB \cite{Samain2013, Riedel2007, Ware2008}) in the chemical formula NH$_4\{$Fe(III)[Fe(II)(CN)$_6]\}$ $\cdot$ x H$_2$O). Iron and trace elements contents were determined by ICP-OES. Carbon, nitrogen and hydrogen were analyzed by CHNS analysis. The results are presented as mass fractions $\omega_{\mathrm{rel}}$ and are averaged values of the replicates. The bias values are the standard deviations.}\label{tab2}

\begin{tabular}{lcc}	
\toprule		
Element & $\omega_{\mathrm{rel}}$ [\%] & $\omega_{\mathrm{rel}}$ [\%] theoretical\\
\midrule
Fe  &    $30.90\pm0.45$ & x = 0: 39.07, 1: 36.76, 2: 34.70, 3: 32.86, 4: 31.21, 5: 29.71 \\
\cr
Traces\textsuperscript{[a]} &&\\ 
\hspace{5pt} total & $0.52\pm0.03$ & \\
\hspace{5pt} K & $0.22\pm0.02$ & \\
\cr
N 	&	 $28.35\pm0.35$ & x = 0: 34.30, 1: 32.27, 2: 30.46, 3: 28.85, 4: 27.40, 5: 26.08\\
\cr
C   &	 $18.43\pm0.11$	& x = 0: 25.21, 1: 23.72, 2: 22.39, 3: 21.20, 4: 20.14, 5: 19.17\\
\cr
H   &	 $2.34\pm0.01$ & x = 0: 1.41, 1: 1.99, 2: 2.50, 3: 2.97, 4: 3.38, 5: 3.75 \\
\bottomrule	
	\end{tabular}
\footnotesize{\textsf{[a] For Mn, Ca, Na, Cr, Cu, S, Zn, Ni, Mg, Ba, Co individual results see SI Table S1.}}

\end{table}

\subsubsection*{Inductively coupled plasma optical emission spectroscopy} In the ICP-OES analysis, the focus was especially on the determination of the iron content. On average an iron mass content of $\omega$(Fe)$_{\mathrm{rel}}$ $=$ 30.90$\%$ $\pm$ 0.45$\%$ was found (see Table \ref{tab2} and S1). Compared to the ideal theoretical value of pure NH$_4$\{Fe(III)[Fe(II)(CN)$_6$]\} ($\omega$(Fe)$_{\mathrm{rel}}$ $\approx$ 39.1$\%$, see Table \ref{tab2} and SI Table S1) the iron content determined differs about 8.20\,pp (percentage points) and is therefore clearly lower, even when considering the mentioned bias of the quantification result. Similar to the case of classic "soluble" potassium PB, the potential crystal water content of the AFCF sample can be assumed as the major reason for the difference. A significant high contamination with the byproduct ammonium chloride, which would also yield to a reduced sample iron content, could be excluded due to the purification procedure as well as the species analyses (see below, for example, XRD results). Thus, the purified AFCF sample examined in this study is a hydrate and the chemical formula actually is NH$_4\{$Fe(III)[Fe(II)(CN)$_6]\}$ $\cdot$ x H$_2$O. A comparison with the theoretical iron content of AFCF with water contents of x $=$ 1 - 5 (the typical range of x in "soluble" PB as described in the literature \cite{Samain2013, Riedel2007, Ware2008}, see Introduction and Table \ref{tab2} and SI Table S1) indicated or, at least, suggested a water content of x $\approx$ 4 in the sample. However, it must be noted that also a potential presence of iron defects (vacancies) in the solid state structure of the sample could cause a lower iron content. Since further indications for defects were found in the CHNS and the Mössbauer analyses (see below) an additional small contribution of these defects to the observed lower iron content is probable.   

Besides iron, several other elements (e.g., K, Mn, Co, and Ni, see Table \ref{tab2} and Table S1) have also been considered in the ICP-OES analyses to check the purity of the AFCF sample in more detail. The contents of these elements were in the range of $\omega$ $\approx$ 20 - 2000\,mg/kg and therefore the general purity of the sample with respect to the trace element contents was confirmed. Nevertheless, the detected potassium content of $\omega$(K) $=$ 2216\,mg/kg $\pm$ 119\,mg/kg must be mentioned. This could originate from the applied educt of ammonium hexacyanoferrate(II), since this compound is usually derived from the potassium salt and therefore potassium hexacyanoferrate(II) \cite{GIESE1988363}. Therefore, potassium residues are a potential impurity of ammonium hexacyanoferrate(II). Hence, the AFCF sample contained potassium to a small extent or, in general, alkali metal cations (when also considering the determined sodium content $\omega$(Na) $=$ 427\,mg/kg $\pm$ 11\,mg/kg, see Table S1) instead of ammonium. It can be assumed that the small amount of potassium cations K$^+$ exchanges ammonium $\mathrm{NH}_4^+$ \cite{Kraft2022, Janiak+2007+381+580} in or directly after PBC formation in the AFCF synthesis process because the effectiveness in alkali binding scales from left to right
\begin{equation}
\mathrm{Na}^+<\mathrm{NH}_4^+<\mathrm{K}^+<\mathrm{Rb}^+<\mathrm{Cs}^+
\end{equation}
as a function of the ionic radii \cite{GIESE1988363,doi:10.4491/eer.2018.177}. In some cases the binding capacity of Cs$^+$ is even increased, because in addition to the described ion exchange mechanism other aspects, such as ion exchange with hydronium ions (simplified H$^+$, from H$_2$O in the PB structure), adsorption and ion trapping can also contribute to the binding process, although these additional aspects will more likely be relevant for "insoluble" PBs \cite{MARTINEZALONSO2024, YANG2008, Altagracia-Martínez2012} and not for PBs "soluble" such as the AFCF sample.

\subsubsection*{Elementary analysis (CHNS analysis)} The focus of the CHNS analysis was on the determination of the nitrogen, carbon and hydrogen contents. In context of a possible sample impurity, sulfur was also considered in the measurements but the concentration was below the limit of detection of this method (in contrast to the ICP-OES measurements, where it was successfully determined, see SI Table S1). 

The obtained results of $\omega$(N)$_{\mathrm{rel}}$ $=$ 28.35$\%$ $\pm$ 0.35$\%$, $\omega$(C)$_{\mathrm{rel}}$ $=$ 18.43$\%$ $\pm$ 0.11$\%$ and $\omega$(H)$_{\mathrm{rel}}$ $=$ 2.34$\%$ $\pm$ 0.01$\%$ differ significantly from the theoretical contents of pure NH$_4\{$Fe(III)[Fe(II)(CN)$_6]\}$ (see Table \ref{tab2} and SI Table S1). In detail, the nitrogen and carbon contents are about 6.0\,pp (N) and 6.80\,pp (C) less whereas the hydrogen content is about 0.93\,pp (H) higher than the theoretical values.

As a comparison with theoretical element contents of AFCF with different water contents x demonstrates (see Table \ref{tab2} and Table S1), reduced N and C in combination with increased H contents are a characteristic evidence for the presence of an AFCF hydrate. Therefore, the CHNS results were in accordance with the ICP-OES results regarding the finding, that the examined AFCF sample is a hydrate of the chemical formula NH$_4\{$Fe(III)[Fe(II)(CN)$_6]\}$ $\cdot$ x H$_2$O. However, the estimated water content x differs from the ICP-OES results and depends on the selected element (see Table \ref{tab2} and SI Table S1). When just taking the determined nitrogen content into account, a water content of x $\approx$ 3 - 4 can be estimated, which is still close to the result based on the ICP-OES iron analysis. In contrast, the hydrogen content, which should be in a particularly close relation to the water content, would indicate lower (x $\approx$ 1 - 2) whereas the carbon content would yield to significantly higher (x $>$ 5) water contents. Although an additional thermogravimetric analysis (TGA) could complement the presented results, it must be pointed out that similar difficulties in the determination of the exact water contents of PB pigments are generally known and described in the literature, e.g., by Samain \textit{et al.} 2013 \cite{Samain2013} and Martinez-Alonso \textit{et al.} 2024 \cite{MARTINEZALONSO2024}.

However, a closer look to the determined element contents and especially to the nitrogen and carbon values reveals another difference to the theoretical or ideal contents. As it can be calculated from the presented values in Table \ref{tab2} (and Table S1), in ideal and pure NH$_4\{$Fe(III)[Fe(II)(CN)$_6]\}$ $\cdot$ x H$_2$O the ratio of the nitrogen and carbon mass contents ($\omega$(N)/$\omega$(C)) is usually about 1.361. In contrast, the ratio of the investigated sample is about 1.539 and, thus, distinctively higher. In detail, the nitrogen content is increased and the carbon content decreased when compared to an ideal and pure AFCF. There are several possible reasons for this difference and two of them could be impurities in the form of significant ammonium chloride admixtures (this would increase the sample nitrogen concentration) or deviations from the ideal AFCF composition and structure in context of, \textit{inter alia}, defects (this would also alter the nitrogen and carbon contents). Similar observations can be conducted regarding the ratios of the iron and carbon as well as nitrogen mass contents ($\omega$(Fe)/$\omega$(C) - ideal: $\approx$ 1.550, sample results: $\approx$ 1.677 and $\omega$(Fe)/$\omega$(N) - ideal: $\approx$ 1.139, sample results: $\approx$ 1.090). As it is the case with the ICP-OES evaluation, major ammonium chloride or other by-product and educt contamination can be excluded and, thus, the latter reason (defects) is assumed.

\subsection*{X-ray powder diffraction}
\begin{figure}
\centering
\includegraphics[width=\textwidth]{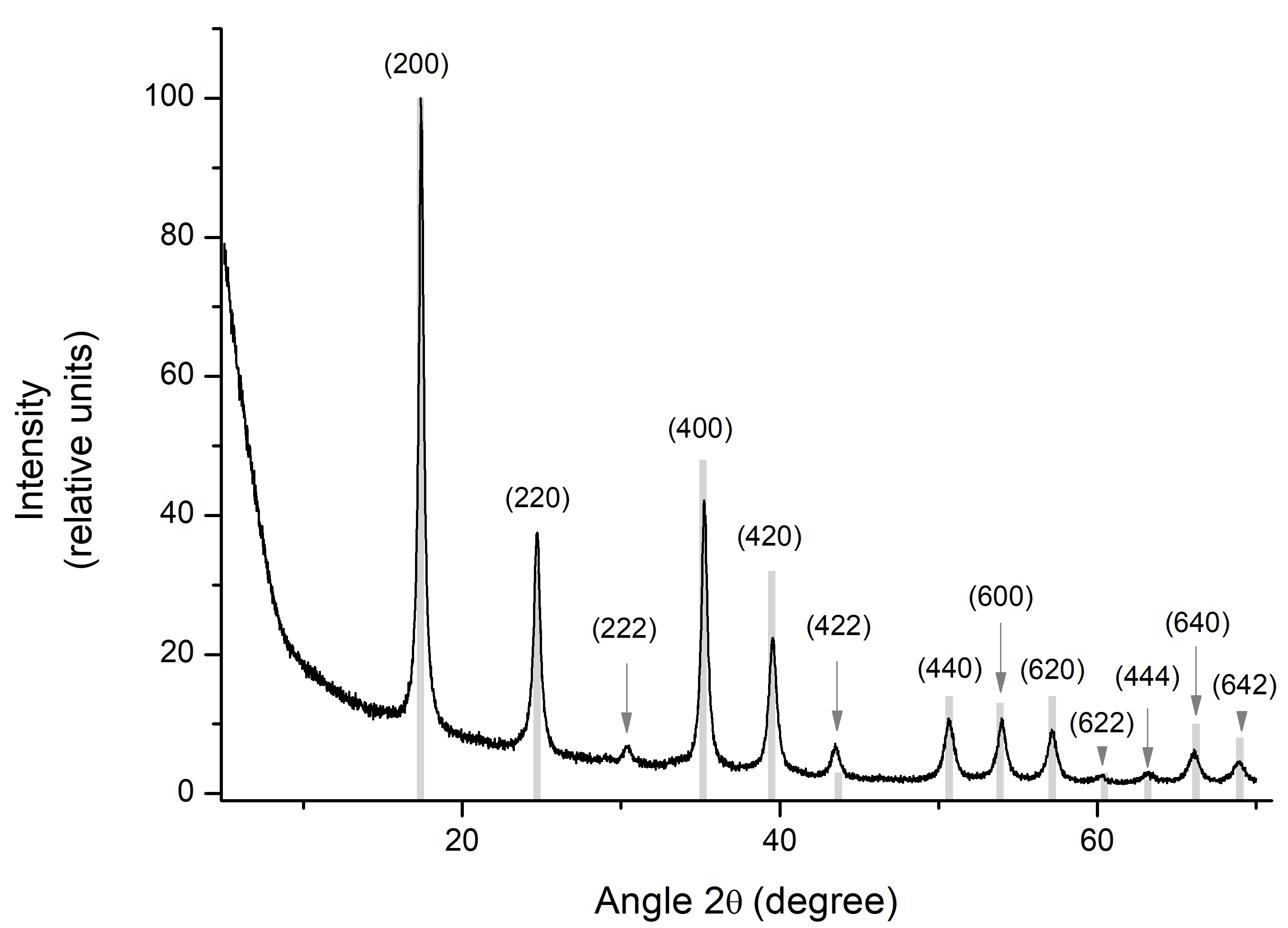}
\caption{Normalized X-ray powder diffractogram of the AFCF sample (radiation: Cu K$\alpha$). The obtained diffractogram is shown in comparison to a reference pattern of an "insoluble" PB (PDF [01-0239], related to Hanawalt \textit{et al.} 1938 \cite{hanawalt}). Additionally, an indexing of the measured reflections has been performed.The AFCF sample shows reflections as expected for a compound exhibiting a Prussian blue structure.}\label{figXRD}
\end{figure}
In Fig. \ref{figXRD} the obtained sample X-ray diffractogram is presented including the accomplished phase analysis and indexing. In total thirteen reflections at 2$\theta$ $=$ 17.40, 24.74, 30.50, 35.22, 39.56, 43.47, 50.64, 53.98, 57.12, 60.29, 63.27, 66.12 and 68.97° were found, which were assigned to the lattice planes (hkl) $=$ (200), (220), (222), (400), (420), (422), (440), (600), (620), (622), (444), (640) and (642) (see also SI Table S2). With regard to signal positions and relative intensities, it is evident that the measured AFCF sample pattern can be explained by the selected "insoluble" Prussian blue reference diffractogram (PDF [01-0239]) virtually completely. As it can be demonstrated by comparing XRDs presented in literature, e.g., Samain \textit{et al.} 2013 \cite{Samain2013}, Carinato \textit{et al.} 2020 \cite{Carniato2020}, Manabe \textit{et al.} 2020 \cite{Manabe2020} and Martinez-Alonso \textit{et al.} 2024 \cite{MARTINEZALONSO2024}, in general, XRD patterns of different "soluble" and "insoluble" PB pigments are known to be very similar, since the basic crystal system, Bravais lattice and crystal structure (simplified a cubic face-centered structure, Fm$\bar{3}$m, see Fig. \ref{figGie}) is identical. Therefore, the AFCF sample of this study can clearly be identified as a PB pigment by applying the "insoluble" PB reference diffractogram. It must be emphasized that the weak reflection unassigned at 30.50 2$\theta$ does not indicate an impurity of the sample, although it does not appear in the reference diffractogram available in this study. In fact, the additional indexing performed of the reflections (see Fig. \ref{figXRD} and SI Table S2) and a comparison to the literature \cite{Samain2013, Carniato2020, Manabe2020, Bueno2008} proved that the signal is a potential part of a PB XRD pattern. Hence, the AFCF sample purity and especially the absence of potential high contents of the reaction by-product ammonium chloride (NH$_4$Cl), educt residues (FeCl$_3$ $\cdot$ 6 H$_2$O, (NH$_4$)$_4$[Fe(II)(CN)$_6$]) or educt-related residues (e.g., (NH$_4$)$_3$[Fe(III)(CN)$_6$] due to oxidation of (NH$_4$)$_4$[Fe(II)(CN)$_6$] by air contact and Fe(III) oxo- / hydroxo-species due to FeCl$_3$ hydrolysis) is evidenced. However, the obtained confined limit of detection (LOD) followed by the proportionally high background and the moderate signal-to-noise ratio \textit{S}/\textit{N} of the applied measurement technique must be considered (see below). 

Furthermore, in Fig. \ref{figXRD} a slight broadening of the signals is also recognizable. On the one hand, reflection broadening is caused by small sample particle sizes or rather crystal sizes (more specifically ordered domains) and indicates that the sample particles are in a nanoscale size, thus, in the range \textit{d} $<$ 150 - 200 nm. However, the particles must also be in the range \textit{d} $>$ 3 - 15\,nm, since is the limit of the transition to X-ray amorphous substances \cite{spiess2009moderne}. In this study, the result underlines the assignment of the AFCF to "soluble" PB pigments. Therefore, an approximation of the crystal sizes of the AFCF sample was made based on the application of the Scherrer FWHM method (see SI equation (S2) and Table S3) \cite{spiess2009moderne, Sharma2012, Ameh2019}. As a result, a size of \textit{D} $\approx$  19.1\,nm $\pm$ 2.3\,nm was obtained (see SI Table S3 for further details), even though a relatively high "real" or more precisely systematic bias must be supposed since the contribution of instrumental broadening to the total reflection broadening could not considered in this study (see Experimental). It has to be underlined that this result is not in opposition to the light microscopy, SEM and LD results and those reported by Giese 1988 \cite{GIESE1988363}. The reason is that with the microscopy and LD in this study as well as with light scattering used by Giese 1988 \cite{GIESE1988363} the sizes of particles still consisting of agglomerates of multiple nanosized grains or crystals have been determined. In rare cases the magnitudes of crystals, grains, and effective particles and, thus, crystal sizes, grain sizes, and particle sizes are equal or, at least, close to each other. However, reflection broadening can also be caused by an increased amount of crystal defects \cite{spiess2009moderne}. Since such defects are of distinct importance regarding PBCs and probably also the AFCF sample of this study (see elemental analysis), it can be assumed that both effects (naturally, always in combination with the instrumental broadening effect as the third effect) yielded the observed total signal broadening, which limits the validity of the Scherrer crystallite size approximation results. Nevertheless, the results were in good accordance with those obtained by the Mössbauer spectroscopic examinations (see below). 

Finally, as a third aspect that influences the quality of the diffractogram, the application of copper K$\mathrm{\alpha}$ radiation ($\lambda$ $=$ 1.5406\,{\AA}) for the XRD measurements  must be mentioned. Although Cu K$\alpha$ radiation is one of the most common radiations in XRD \cite{spiess2009moderne, Mos2018}, it is generally known to have limited suitability for XRD analysis of iron compound samples or iron compound containing samples because high absorption and induced Fe X-ray fluorescence can decrease the signal intensities of the reflections, increase the background, and worsen the \textit{S}/\textit{N} ratios \cite{Mos2018}. Depending on the diffractogram quality required to deal with a specific iron-containing sample and the related individual scientific analytical question or task as well as potentially available XRD setup modifications (e.g. monochromators), which can reduce the negative impact mentioned of Cu K$\mathrm{\alpha}$ radiation application, especially the use of cobalt K$\mathrm{\alpha}$ radiation ($\lambda$ $=$ 1.790\,{\AA}) can become vital \cite{Mos2018}. Thus, it can be expected that the application of Co K$\mathrm{\alpha}$ radiation could produce an improved AFCF sample diffractogram and, hence, improved results regarding the examination of sample purity and crystallite sizes. 

\subsection*{Vibrational spectroscopy}

\begin{figure}%
\centering
\includegraphics[width=\textwidth]{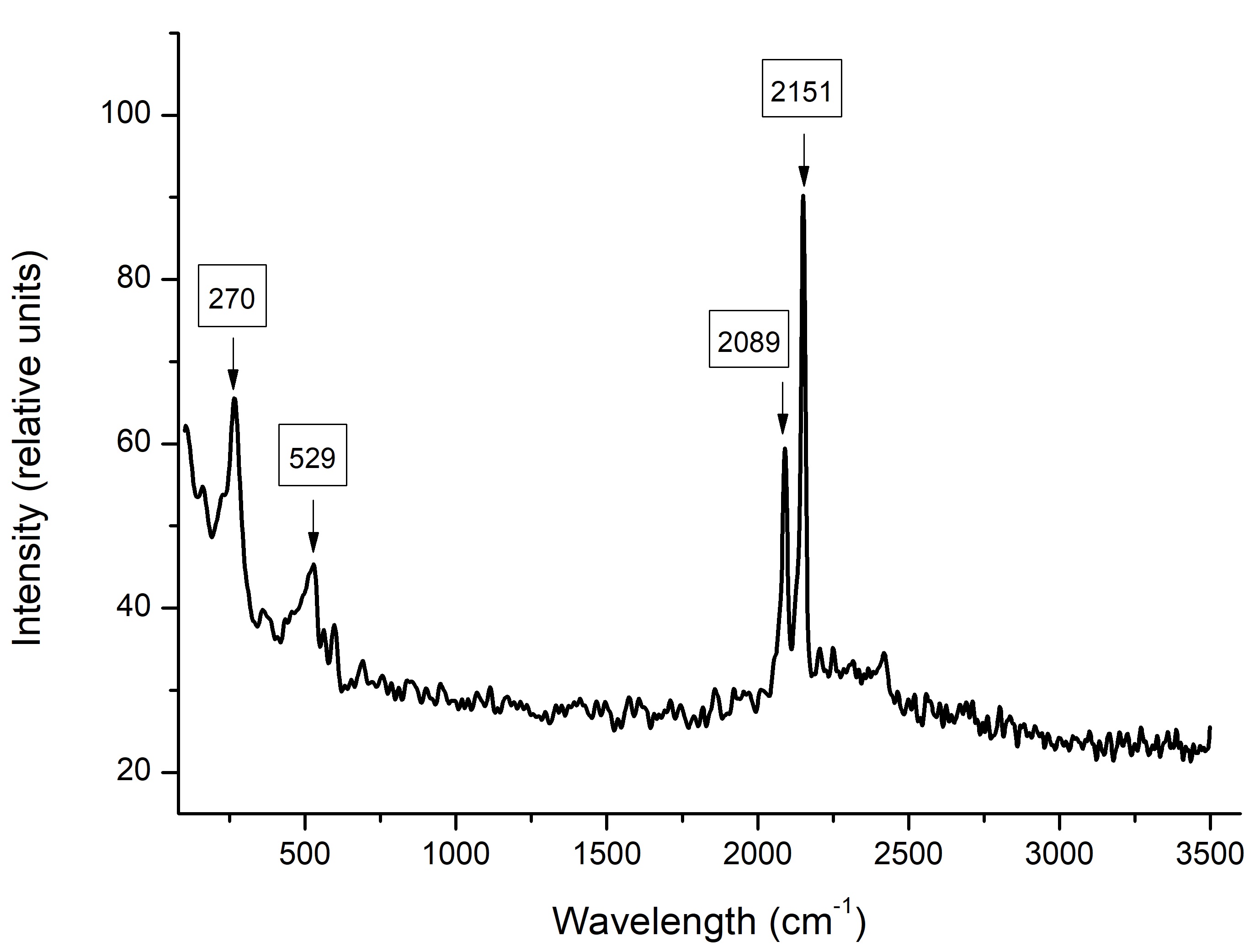}
\caption{Smoothed Raman spectrum of the AFCF sample at $T\approx300$\,K. The wave number positions of the observed most essential bands are marked. The obtained excitation frequencies confirm typical C$\equiv$N-valence, Fe$-$C$\equiv$N$-$Fe-deformation and Fe$-$C-stretching vibration modes as discussed in PB by G. Mooretti \textit{et al.} \cite{Moretti2018RamanSO}.}\label{figRaman}
\end{figure}
\begin{table*}
\caption{Dominant Raman wavenumbers as presented in Fig. \ref{figRaman} (rounded to natural numbers, an error of $\pm3$\,cm$^{-1}$ is concluded by systematic and statistical deviations). Errors of the values of the literature are unknown.}\label{tab1}
\begin{tabular}{lcc}
\toprule
Raman frequency (cm$^{-1}$) & Literature \cite{Moretti2018RamanSO}&  Identified vibration mode  \\
\midrule
270  & 276 & Fe$-$C$\equiv$N$-$Fe-Deformation vibration \cite{Moretti2018RamanSO}   \\
529  & 510 & Fe$-$C stretching mode \cite{Moretti2018RamanSO}     \\
2089 &  2091 & C$\equiv$N valence mode  \cite{Moretti2018RamanSO}   \\
2151  & 2152 & C$\equiv$N valence mode \cite{Moretti2018RamanSO}     \\
\bottomrule	
\end{tabular}
\end{table*}
PBCs are pigments highly sensitive to electromagnetic radiation because of the complex redox behavior, e.g., similar sensitivity is observed in phthalocyanine pigments \cite{ramanlib}. From the physicochemical perspective, Raman and IR spectroscopy are complementary techniques regarding the structural information obtained from the spectra because with IR spectroscopy time dependence of the electric dipol moment and with Raman spectroscopy time dependence of the polarization are measured. Both techniques were applied to the AFCF sample in this study.
Fortunately, ammonium is a molecule, and an additional excitation band for NH$_4^+$ modes is expected. Apart from that, the main network is similar according to the crystal structure and local coordination discussed above by the XRD results. Therefore, it is reasonable to compare the spectral signature of Prussian Blue and \textit{Giese-Salt} focusing on the main network structure. 
\subsubsection*{Raman spectroscopy}
Fig. \ref{figRaman} shows the of Raman spectrum up to 3500\,cm$^{-1}$ of the Stokes branch and in Table \ref{tab1} the Raman frequencies are compared to the literature of Prussian blue \cite{Moretti2018RamanSO}. The four dominant vibration modes are clearly identified and fully consistent with the associated Raman intensities of the network structure \cite{Moretti2018RamanSO}. A slight shift is explained by variability of the laser power, excitation frequency and a chemical variability of the host network as observed by XAFS and Mössbauer spectroscopy, see below. However, there is one point that should be emphasized in Table \ref{tab1}. The literature value of the Fe$-$C stretching mode should be understood of value with the maximal deviation according to Moretti \textit{et al.} \cite{Moretti2018RamanSO}. In case of "soluble" Prussian blue, the dominant resonance by intensity is given by the wavenumber around 510\,cm$^{-1}$, and 529\,cm$^{-1}$ is the observed band in ammonium Prussian blue. However, in "insoluble" Prussian blue the intensity of two peaks around 528\,cm$^{-1}$ and 509\,cm$^{-1}$ is inverted \cite{MAZEI2011}. In general, the bands around 598, 528, 509 and 445\,cm$^{-1}$ are considered to stretching vibrations of Fe–C molecule fragments \cite{MAZEI2011}. High-pressure studies confirming the main value of 511\,cm$^{-1}$ \cite{barsam2011} at zero-pressure, which means ambient pressure. It may be that these modes can be understood as probes for defects and alkali ions, which explains their variance possibly based on various synthesis methods using electrochemistry  \cite{MAZEI2011}. In any case, XAFS spectroscopy measured a change in bond length (see below) that would also explain a deviation by a literature value of 510\,cm$^{-1}$ according to Moretti \textit{et al.} \cite{Moretti2018RamanSO}.

\subsubsection*{Infrared spectroscopy} By applying ATR-IR spectroscopy to the AFCF sample in total eight bands (see obtained ATR-IR spectrum SI Fig. S7 and Table S4) with the wave numbers 600\,cm$^{-1}$, 829\,cm$^{-1}$, 981\,cm$^{-1}$, 1410\,cm$^{-1}$, 1600\,cm$^{-1}$, 2060\,cm$^{-1}$, between 2600 and 3500\,cm$^{-1}$ (very broad) and 3230\,cm$^{-1}$ have been identified. A comparison to the IR spectra of PB pigments presented or described in the literature \cite{Monteiro2017, Penche2022, NAWAR2023, Vahur2016} as well as to the spectra of the synthesized reference PBs of this study (these were pre-characterized with, \textit{inter alia}, ATR-IR before their application as reference substances in the XAFS examinations, see below and SI) reveals a distinct accordance with the typical vibration bands of PB pigments regarding band positions, shapes, and intensities (see SI Table S4). 

Thus, analogously to the Raman spectrum, the characteristic network of Fe(II)$-$C$\equiv$N$-$Fe(III) reflected in the measured IR spectrum. The detected bands at 1600\,cm$^{-1}$ and 2600 - 3500\,cm$^{-1}$ have to be highlighted because they can be assigned to the bending $\delta$(H-O-H) and stretching vibration $\nu$(OH) \cite{Monteiro2017, NAWAR2023}. Therefore, these bands further confirm the presence of water and, therefore, the conclusion of the elemental analysis (see above) of the identity of the AFCF sample as a hydrate of the type NH$_4\{$Fe(III)[Fe(II)(CN)$_6]\}$ $\cdot$ x H$_2$O. Furthermore, the observed bands at 1410\,cm$^{-1}$ and 3230\,cm$^{-1}$ must be emphasized. These bands, which are absent in the reference PBs spectra as expected, can clearly be assigned to bending $\delta$(N-H) and stretching vibration $\nu$(NH) when comparing the AFCF spectra additional to literature spectra of different ammonium compounds (e.g., NH$_4$Cl) \cite{Miller1952, Hesse2002}, Thus, the presence of these two bands is a final distinct evidence for the presence of ammonium (NH$_4^+$) in the sample.  

\subsection*{UV-Vis/NIR spectroscopy}
In the UV-Vis/NIR spectrum obtained from the AFCF sample dispersed in (acidic) aqueous medium (depicted in SI Fig. S8) four essential features can be identified. In general, the characteristics of these four spectra fit well to the numerous spectra of PB pigments presented or described in the literature \cite{Samain2013, Agrisuelas_2009, Monteiro2017, Onoe2021, Melvin1962, Watanabe2016} and to the physical background of the interaction of PB pigments with electromagnetic radiation. 

Firstly, a very strong and broad absorption band in the visible and near-infrared range (band range $\lambda$ $\approx$ 510 - 1300\,nm, see SI Fig. S8) with a maximum at $\lambda$$_{max1}$ $=$ 722.5\,nm $\pm$ 0.5\,nm was observed. This band can mainly be assigned to the intervalence charge-transfer (IVCT or IT) between the CN-bridged Fe(II) and Fe(III) (Fe(II)$_{\mathrm{LS}}$$-$C$\equiv$N$-$Fe(III)$_{\mathrm{HS}}$), in detail, the transition (Fe$_\mathrm{C}$: t$_{2\mathrm{g}}^6$e$_\mathrm{g}^0$ or $^1$A)(Fe$_\mathrm{N}$: t$_{2\mathrm{g}}^3$e$_\mathrm{g}^2$ or $^6$A) $\rightarrow$ (Fe$_\mathrm{C}$: t$_{2\mathrm{g}}^5$e$_\mathrm{g}^0$ or $^2$T)(Fe$_\mathrm{N}$: t$_{2\mathrm{g}}^4$e$_\mathrm{g}^2$ or $^5$T), which technically switches the oxidation states of the two iron cations (Fe(III)$_{\mathrm{LS}}$$-$C$\equiv$N$-$Fe(II)$_{\mathrm{HS}}$) \cite{Samain2013, Agrisuelas_2009, Monteiro2017, Onoe2021, Melvin1962, Watanabe2016}. Since CT transitions are both spin- and Laporte-allowed and, hence, have a high transition probability (high cross sections), they usually cause absorption bands of high intensities or rather high absorbance \cite{Riedel2007, Janiak+2007+381+580, Melvin1962}, as is the case with the AFCF sample. However, it should be mentioned that the additional participation of a two-electron IVCT ((Fe$_\mathrm{C}$: t$_{2\mathrm{g}}^6$e$_\mathrm{g}^0$ or $^1$A)(Fe$_\mathrm{N}$: t$_{2\mathrm{g}}^3$e$_\mathrm{g}^2$ or $^6$A) $\rightarrow$ (Fe$_\mathrm{C}$: t$_{2\mathrm{g}}^4$e$_\mathrm{g}^0$ or $^3$T)(Fe$_\mathrm{N}$: t$_{2\mathrm{g}}^5$e$_\mathrm{g}^2$ or $^4$T)), 3d - 3d transitions after or in combination with the described IVCT ((Fe$_\mathrm{C}$: t$_{2\mathrm{g}}^5$e$_\mathrm{g}^0$ or $^2$T)(Fe$_\mathrm{N}$: t$_{2\mathrm{g}}^4$e$_\mathrm{g}^2$ or $^5$T) $\rightarrow$ (Fe$_\mathrm{C}$: t$_{2\mathrm{g}}^5$e$_\mathrm{g}^0$ or $^2$T)(Fe$_\mathrm{N}$: t$_{2\mathrm{g}}^3$e$_\mathrm{g}^3$ or $^5$E) or (Fe$_\mathrm{C}$: t$_{2\mathrm{g}}^6$e$_\mathrm{g}^0$ or $^1$A)(Fe$_\mathrm{N}$: t$_{2\mathrm{g}}^3$e$_\mathrm{g}^2$ or $^6$A) $\rightarrow$ (Fe$_\mathrm{C}$: t$_{2\mathrm{g}}^5$e$_\mathrm{g}^0$ or $^2$T)(Fe$_\mathrm{N}$: t$_{2\mathrm{g}}^3$e$_\mathrm{g}^3$ or $^5$E)) as well as different spin-transitions of the Fe(III) on this band are also suggested \cite{Melvin1962, Watanabe2016}. The observed mentioned band range and the position of the maximum of the examined AFCF sample are also in good agreement with the literature, where values of $\lambda$$_{max}$ $\approx$ 680 - 730\,nm \cite{Samain2013, Monteiro2017}, depending on the specific pigment composition of PB ("soluble" or "insoluble" PB, presence and type of alkali metal cations or ammonium cations, purity), particle size, and applied solvent (solvent type, pH value, etc.), have been reported. 

This CT band is followed by a slight shoulder (see arrow in SI Fig. S8) in the transition zone between the visible and UV range at $\lambda$ $\approx$ 400\,nm. This feature is primary caused by spin-allowed 3d - 3d transitions of the LS-Fe(II), e.g. the transition (Fe$_\mathrm{C}$: t$_{2\mathrm{g}}^6$e$_\mathrm{g}^0$ or $^1$A) $\rightarrow$ (Fe$_\mathrm{C}$: t$_{2\mathrm{g}}^5$e$_\mathrm{g}^1$ or $^1$T) \cite{Monteiro2017, Watanabe2016}. The relatively weak intensity of this band is a consequence of the general lower probability (low cross sections) of the d - d transitions (especially in the case of octahedral coordination) because these are Laporte forbidden \cite{Riedel2007, Janiak+2007+381+580}. The observed position of this band in the spectra of the AFCF sample is in perfect accordance with the literature \cite{Monteiro2017}, although it must be mentioned that it was obtained in the form of a barely resolved weak shoulder instead of a separated weak band. 

In the UV range two narrower, overlapping and moderately intensive bands (in comparison to the strong CT and weak 3d - 3d band) with absorption maximums at $\lambda$$_{max2}$ $=$ 322.5\,nm $\pm$ 0.5\,nm as well as $\lambda$$_{max3}$ $=$ 276.0\,nm $\pm$ 0.5\,nm are exhibited. These can also be explained by charge-transfer processes, but in this case between the ligands and the metal cations, the so-called ligand-metal charge-transfer (LMCT) (e.g. from CN$^-$ to Fe$^{2+}$) \cite{Monteiro2017}. The shape and positions determined in the UV range correspond to the description in the literature of the pigment spectra of PB \cite{Monteiro2017, Watanabe2016}.    

\subsection*{Fe K-edge XAFS spectroscopy}
In this section, the results of the obtained AFCF sample, as well as reference substances, are presented and discussed in detail. The theoretical foundation for these descriptions are general XAFS spectroscopy basics and the effects of the chemical and structural environment of the absorber atom (thus, for example, oxidation and spin state, coordination number and polyhedral, bond length or distance) \cite{Bunker_2010, Schnohr2015, Kas2016, Joly2016} as well as general descriptions of XAFS spectra at the K-edge of iron of several iron compounds \cite{Wilke2001, Matsukawa_1978, PETIAU1988237, Lennie1996, Yamashige2005, Dangelo2008, Motz2023} as described in the literature. Therefore, only specific aspects of PBCs or iron-cyano compounds are individually referenced in the following. 
\begin{figure}%
\centering
\includegraphics[width=\textwidth]{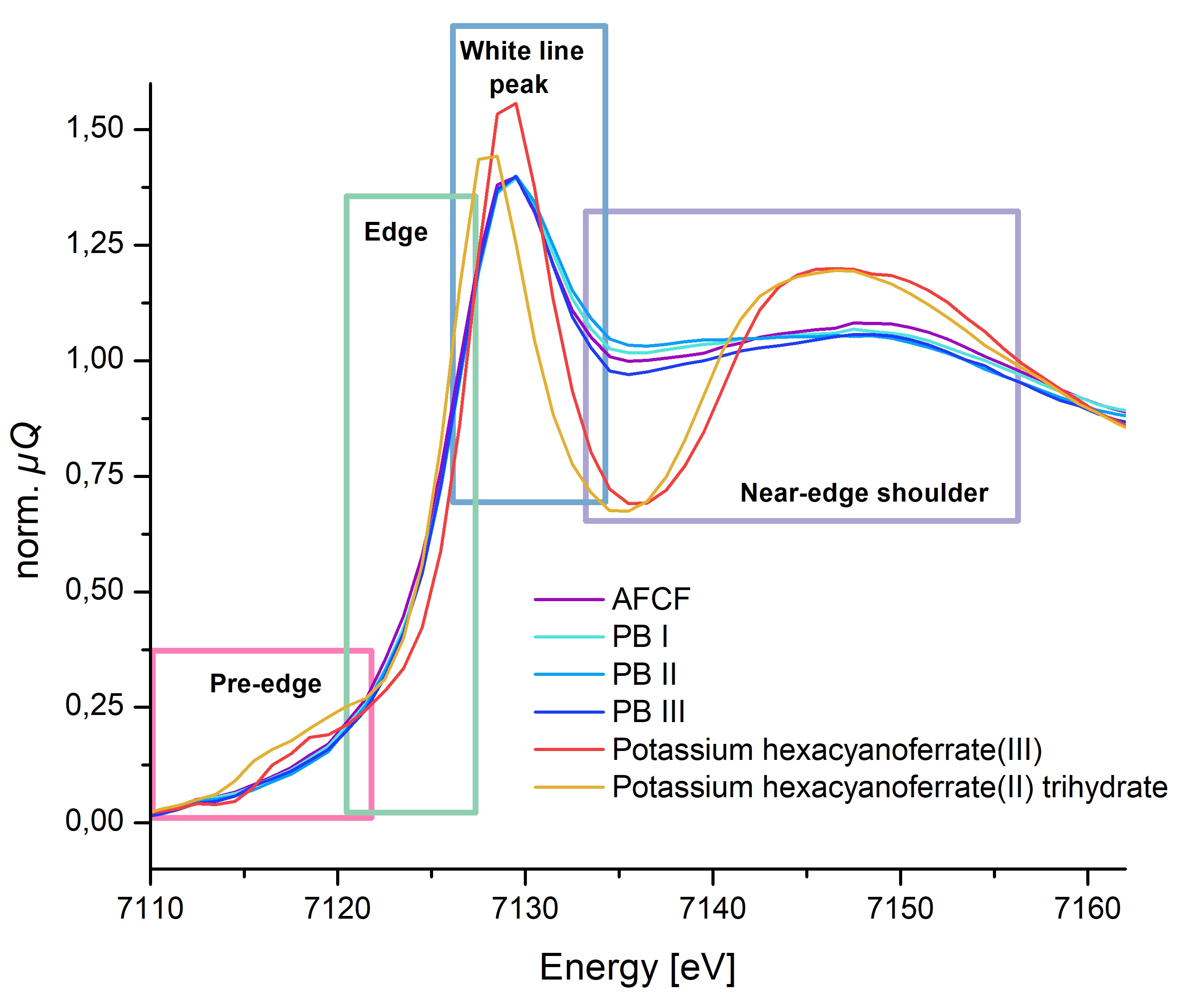}
\caption{Fe K-edge XAFS spectra of the AFCF sample and the reference compounds with the focus on the XANES areas including pre-edges (for an additional better visibility of the PB I, II, III and AFCF pre-edges, see SI Fig. S9b). Important regions of the measured samples, which are discussed in the text, are highlighted. The measurements were performed at room temperature. The complete XAFS spectra (XANES and EXAFS) are additionally depicted in SI Fig. S9a.}\label{fig:xaf}
\end{figure}
In Fig. \ref{fig:xaf} the obtained XAFS spectrum with the focus on the XANES area, including the pre-edges, of the AFCF sample is presented in comparison to the chosen reference substances. A  depiction of the complete XAFS spectra can be seen in the SI (Fig. S9a). In addition, the edge positions have been determined. These are also shown in Fig. \ref{fig:xaf} and the determined values are presented in SI Table S5. 

In general, the spectra of the reference substances K$_4$[Fe(II)(CN)$_6$] ($\cdot$ 3 H$_2$O) and K$_3$[Fe(III)(CN)$_6$] show pronounced similarities regarding the shapes of the spectra in the XANES (including the pre-edge and edge) and EXAFS area. Thus, both reveal a very weak and from the edge just barely separated pre-edge which transitions via a slightly more intensive shoulder into the main edge. The low pre-edge intensities reflect the close to ideal octahedral coordination of the iron(II) and iron(III) by the cyanide ligands since in this case (see electromagnetic selection rules) the 1s - 3d electron transitions are just feasible by electric quadrupole transitions (E2) \cite{Hayakawa2004, Ross2018}. These transitions generally have a lower cross section than electric dipole transitions (E1). The closeness to the main edges trace back to the LS-state of both hexacyanoferrates (Fe(II) ls-3d$^6$: t$_{2\mathrm{g}}^6$e$_\mathrm{g}^0$ or $^1$A, Fe(III) ls-3d$^5$: t$_{2\mathrm{g}}^5$e$_\mathrm{g}^0$ or $^2$T) which is caused by the fact that cyanide is a strong ligand and yields to reduced bond distances and large crystal or ligand field splittings $\Delta$$_O$. Hence, higher energies (closer to the edge energies) for the 1s - 3d are required. In case of the obtained hexacyanoferrate(III) spectrum it is possible to discern faint indications of two individual  features or peaks (a first one at around 7112\,eV and a second one at around 7116\,eV) in the pre-edge region before the edge transition shoulder which can be assigned to the 1s electron transitions into the 3d,t$_{2\mathrm{g}}$ (first peak) and 3d,e$_\mathrm{g}$ (second peak) orbitals \cite{Hayakawa2004, Ross2018}. In contrast, in the obtained hexacyanoferrate(II) spectrum the individual pre-edge feature for the possible 1s electron transition into the 3d,e$_\mathrm{g}$ orbitals as described in the literature \cite{Hayakawa2004, Ross2018} is not clearly visible and not resolved or separated from the transition shoulder. This might be a limitation of the actual resolving power of the applied laboratory XAFS set-up. The slightly more intense shoulders in the transitions zones between pre-edges and edges (at around 7118\,eV for the hexacyanoferrate(III) and around 7116\,eV for the hexacyanoferrate(II)) are caused by transitions with larger cross sections to which also the $\pi$* orbitals of the CN$^-$ contribute \cite{Ross2018}. Therefore, the more detailed molecular orbital theory, instead of the simplified strong-field theory (more specifically crystal-field theory) or weak-field theory, must be considered. The shapes of the main edges, representing the allowed dipole 1s - 4p / np transitions, as well as the following oscillations (scattering effects), are also basically identical due to similar molecular or complex structures of both hexacyanoferrates. Nevertheless, despite these distinct similarities, a shift of the hexacyanoferrate(III) spectra (see pre-edge and edge positions Fig. \ref{fig:xaf} and SI Table S5) of about $\Delta$\textit{E} $\approx$ 0.84\,eV (for the edges) towards higher energies is visible which reflects the differences in the oxidation states and the connected different bond distances. In total, the electronic transitions for the Fe(III) complex compound require more energy than for the Fe(II) compound. However, in case of the two cyano complexes the metal-ligand bond distances are actually smaller for the Fe(II) than for the Fe(III) complex by virtue of $\pi$-backbonding effects occurring in the Fe(II)-CN bonds (in [Fe(II)(CN)$_6$]$^{4-}$: \textit{d}(Fe-C) $\approx$ 1.90 - 1.91\,\AA, in [Fe(III)(CN)$_6$]$^{3-}$: \textit{d}(Fe-C) $\approx$ 1.93 - 1.94\,\AA) \cite{Janiak+2007+381+580, Ross2018}. Hence, it can be assumed that the observed shift primarily traces back just to the direct effect of oxidation state or formal charge.  

The spectra of the reference substances PB I and II (both "insoluble" PBs) as well as PB III ("soluble" PB) exhibit distinct similarities with each other but are different from the hexacyanoferrate spectra. Firstly, the pre-edges and edges are shifted toward lower energy compared to hexacyanoferrate(III), which originates from PB pigments being mixed-valence compounds, and thus containing Fe(III) and Fe(II). It is also visible that there is just a very small edge shift towards higher energies compared to hexacyanoferrate(II), which is also typical for PB pigments (see for example Samain \textit{et al.} 2013 \cite{Samain2013}). Since both "soluble" and "insoluble" PB pigments contain Fe(II) and Fe(III) in a molar ratio close to 1:1 one could expect pre-edge and edge-positions in the middle between the hexacyanoferrate(II) and (III) positions when just considering the contribution of the oxidation states. However, differences in the coordination or bond states must also be considered, which limits the comparability of the spectra of the hexacyanoferrates and PB pigments. Whereas in the hexacyanoferrates CN$^-$ functions as a cyanide ligand and, respectively, a strong ligand for both Fe(II) and Fe(III), in the PB pigments, the anion is a cyanide ligand for Fe(II) but an isocyanide and thus a weak ligand for Fe(III). Therefore, less energy is required for the 1s - 3d (pre-edge) and 1s - 4p / np (edge) transitions and, because the contributions of the different oxidation states (especially to the edge) are inseparable \cite{Samain2013}, the total (pre-) edge energy is closer to that of the hexacyanoferrate(II). In detail the edge positions of all PBs examined in this study are around 7124\,eV which is close to the edge position of hexacyanoferrate(II) at $E$ $=$ 7124\,eV $\pm$ 0.01\,eV (see Table S5). Despite this, the positions of the PB principal maximums (white lines, the edge maximums) are closer to hexacyanoferrate(III). Secondly, the pre-edges of the PB reference substances are of a weak intensity similar or, more precisely, even weaker compared to that for the hexacyanoferrates, but are more separated and, respectively, more distant from the edge (for a better visibility of the measured PBC pre-edges see SI Fig. S9b). Regarding their positions values at around 7110\,eV can be identified in all PB spectra measured in scope of this study (see Fig. S9b). This shows the similar symmetric coordination polyhedral (octahedral for both Fe(II) and Fe(III)) present in the PB structure (similar to the hexacyanoferrates primary just E2 1s - 3d transitions are feasible) but also the mixed spin state (Fe(II) ls-3d$^6$: t$_{2\mathrm{g}}^6$e$_\mathrm{g}^0$ or $^1$A, but Fe(III) hs-3d$^5$: t$_{2\mathrm{g}}^3$e$_\mathrm{g}^2$ or $^6$A) of PB pigments. Finally, the area right after the transition from the white line peak to the near-edge shoulder region (see Fig. \ref{fig:xaf} and SI Fig. S9b) must be emphasized. Regarding XAFS examinations of PBCs this specific area is frequently called slope-region \cite{Samain2013, Gervais2013}. As is apparent from Fig. \ref{fig:xaf} the steepness of this region is different in all three reference substance spectra but reproducible in repetitive measurements of the same pellet and different pellets of a sample (see SI Fig. S9b). In detail the steepness relatively low in the case of PB I and II and higher in the case of PB III. Since the XANES region is primary affected by (multiple) scattering of the photoelectron wave in the vicinity of the absorber atom, it represents the Fe(II)$-$C$\equiv$N$-$Fe(III) system and variations from this ideal system (in the form of distortions, bond distance variations, vacancies \cite{Samain2013, Gervais2013}) between different samples of PB.

In general, the obtained XAFS spectrum of the AFCF sample fits to the PB reference substance spectra distinctively well with respect to the XANES (pre-edge and edge structures or shapes and positions as well as the near post-edge fine structures) and the EXAFS region (fine structures, more specifically oscillations). So far, analogous to the elemental analysis, XRD, vibrational spectroscopy, and UV-vis results, the examined AFCF sample can clearly be identified as a PB pigment with the characteristic structural features. 

However, in the XANES area two differences or, at least, sample-specific characteristics can be identified by comparing the AFCF spectrum with the PB reference substance spectra. Firstly, the steepness of the slope is specific and can be assigned in between the slopes of PB I / II (although the slopes of PB I and PB II are not completely equal, they are very similar at least) and PB III, although it is by trend more similar to the slope development of PB III. This larger similarity to the "soluble" PB III could be interpreted as a further hint that the AFCF sample is also a "soluble" PB pigment, but it must be emphasized and considered that there has been no systematic correlation between the slope steepness and the identification or differentiation between "insoluble" and "soluble" PBs reported in the literature \cite{Samain2013}. The steepness and shape of the slope differ distinctively between individual samples regardless of the "soluble" or "insoluble" stoichiometry \cite{Samain2013}. This is also observable in the results of this study. As has already been mentioned, the slopes of the reference substances PB I and II are indeed very close to each other but not completely equal. Although both substances are classic "insoluble" PBs and basically have the same rough stoichiometry and structure, these differences in the X-ray absorption fine structure occur and reflect slight variances in the structure resulting from differing syntheses conditions. Secondly, there is also a small shift of the AFCF sample edge of $\Delta$\textit{E} $\approx$ 0.27\,eV (see SI Table S5) towards lower energy. This can also be observed for the pre-edge (see SI Fig. S9b).  

\begin{figure}%
\centering
\includegraphics[width=\textwidth]{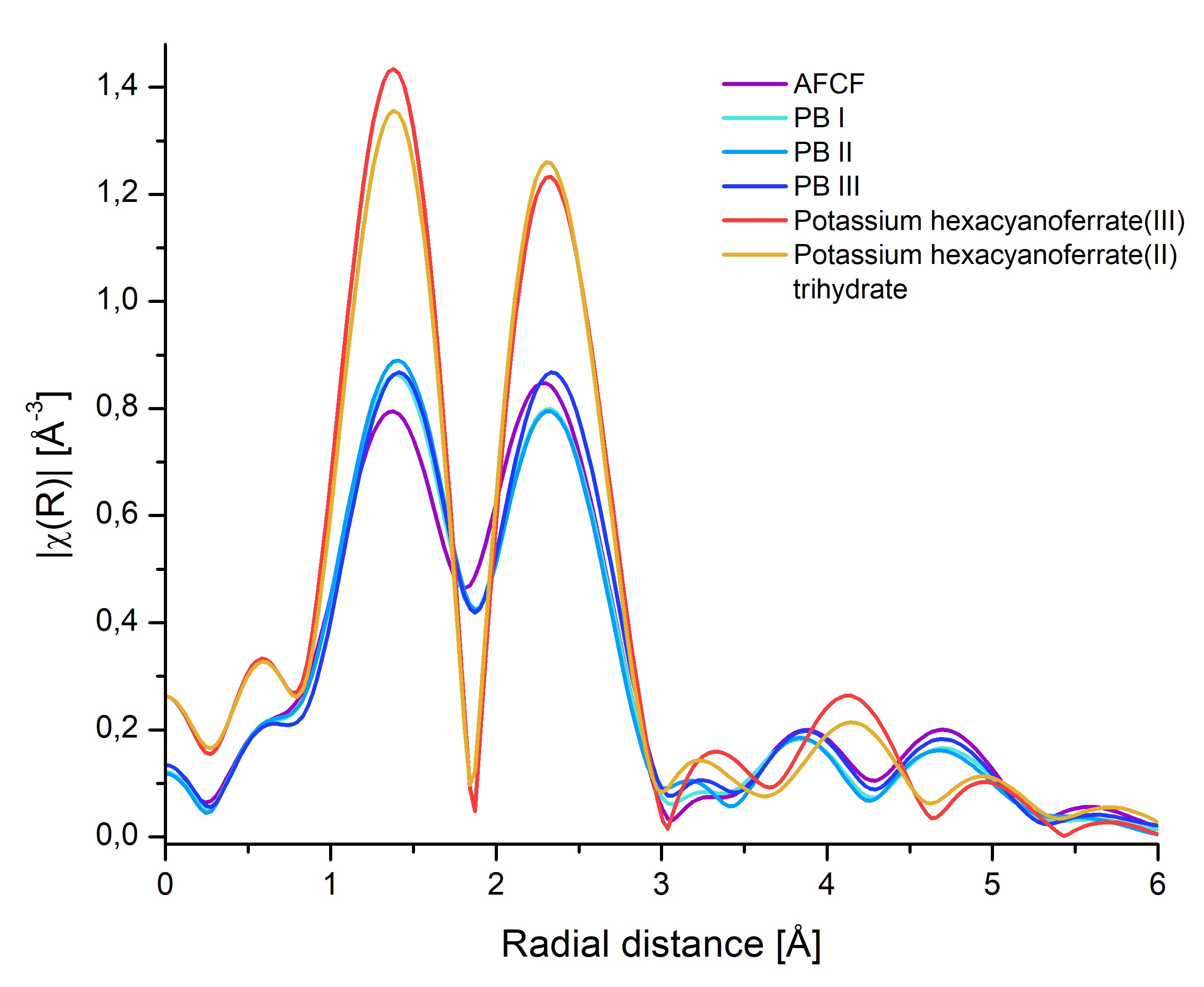}
\caption{FT-EXAFS of the AFCF sample and reference compounds Fe K-edge XAFS spectra. Hence, the EXAFS spectra are presented in the \textit{R}-space. The k-range for the FT was 3 - 8. The measurements were carried out at room temperature.}\label{fig:exafs}
\end{figure}

These observed differences from the reference PBs are further confirmed when the EXAFS data are evaluated. Fig. \ref{fig:exafs} presents the Fourier transform (FT) EXAFS spectra (hence, the EXAFS spectra in the \textit{R}-space) of the reference compound PBs and the AFCF sample. Generally, the oscillation shapes and positions obtained for both the reference substances and the AFCF sample are in good agreement with those reported for PB pigments in the literature \cite{Glatzel2002}. Nevertheless, at lower wave numbers \textit{k} (\textit{k} $<$ 5.5\,{\AA}$^{-1}$) or lower radial distances \textit{R} a shift of the AFCF sample EXAFS towards lower values is clearly visible.

Thus, these spectral differences indicate slight structural differences in the AFCF sample compared to the PB references. Potential reasons, especially for the bathochromic (pre-) edge shift, can be, for instance, a partial reduction of the oxidation state (reduction of Fe(III) to Fe(II)) and / or enlarged bond lengths in the Fe(II)$-$C$\equiv$N$-$Fe(III) network. The first aspect can be excluded for three reasons. Firstly, a (partial) reduction in Fe(III) sites in PB pigments produces significant color fading (basically, Prussian white is the final product of a fully reduced PB) because the IVCT is disturbed \cite{Gervais2013}. This was not observed in either the macroscopic examinations or the UV-Vis spectra. Secondly, in addition to electrochemically induced PB reduction \cite{Kraft2022, Gervais2013}, PB pigment fading can especially be caused by (long-term and direct) UV-Vis light exposure, for example, exposure to sunlight \cite{Samain2013, Gervais2013}. During this study, longer direct exposure of the AFCF sample to sunlight or other UV-Vis light sources was avoided. Third, PB pigment reduction requires a reduction agent (an electron donor), usually in the form of an additional second compound (impurities, a substrate, thus paper in case of paintings, etc.), since it is known that pure PB pigment powders barely fade and show a distinct light fastness \cite{Gervais2013}. As the examined AFCF sample is of proved high purity, the presence of such external reduction agents can also be excluded. Hence, slightly modified, in detail, enlarged bond lengths in the examined AFCF sample seem to be the most probable cause for the determined XAFS shifts.

\subsection*{$^{57}$Fe Mössbauer spectroscopy}
\begin{figure}%
\centering
\includegraphics[width=\textwidth]{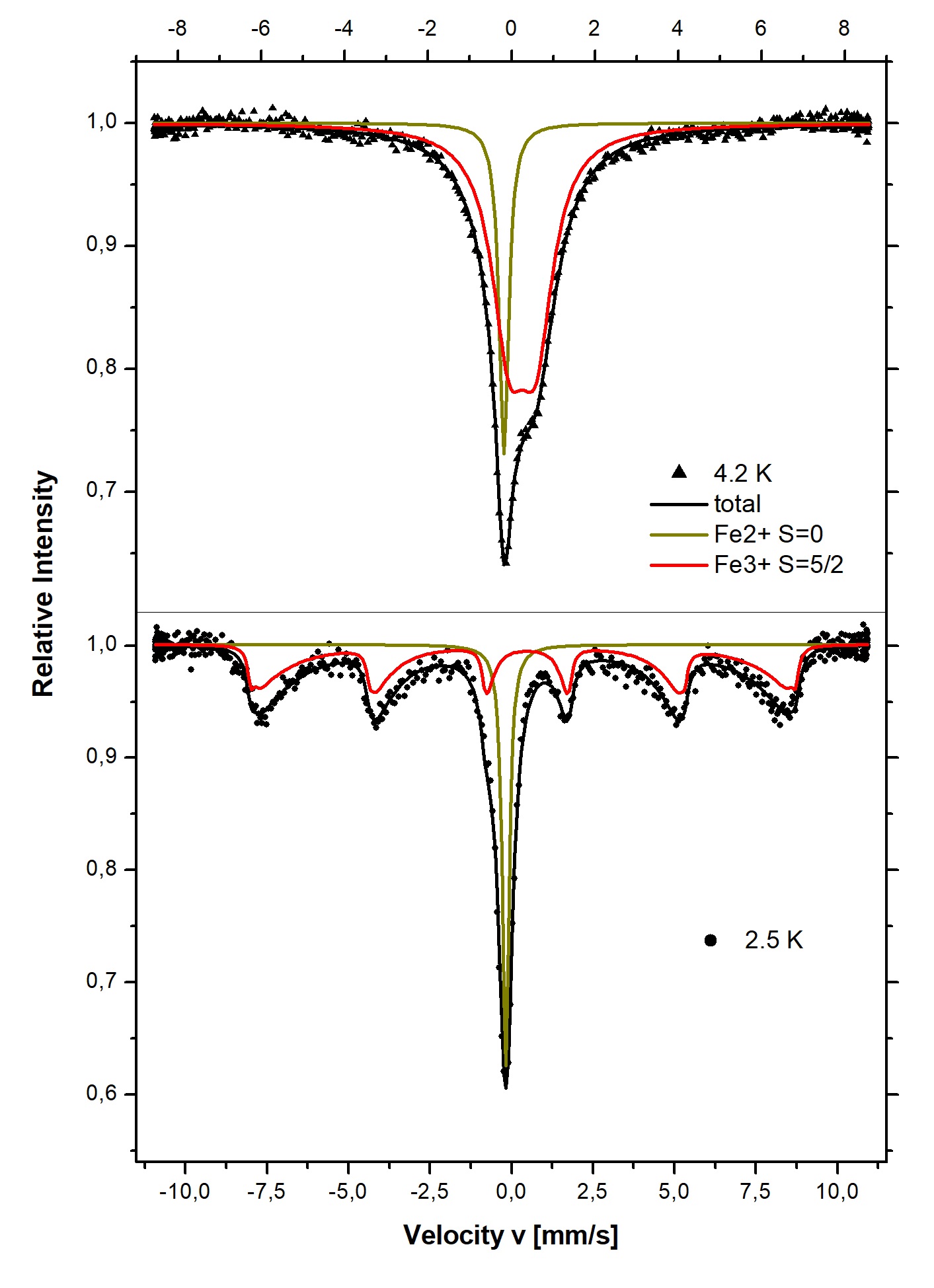}
\caption{$^{57}$Fe Mössbauer spectra of the AFCF sample at 4.2\,K and 2.5\,K. The Fe species are Fe(II) $S=0$ (green) and Fe(III) $S=5/2$ (red).}\label{figmoess}
\end{figure}
Studies have shown that $^{57}$Fe Mössbauer spectroscopy is a strong technique to distinguish between PB species and even to estimate rough upper sizes of spin clusters by magnetic hyperfine interactions in some cases. As has already been stated in the introduction, the hyperfine parameters of PBCs are also known to depend on the way of chemical synthesis by different provenances caused by hypersensitivity \cite{reguera_mossbauer_1992}. To overcome this difficulty of variability in our study, the AFCF sample was synthesized using the same equipment, routine, and conditions as presented by Giese 1988 \cite{GIESE1988363}.
Furthermore, a numerical reinvestigation of historical $^{57}$Fe Mössbauer studies has motivated us to investigate the \textit{Giese-salt} by $^{57}$Fe Mössbauer spectroscopy in a broad temperature range between 2.5\,K and 300\,K\footnote{Interestingly, historical results \cite{Ito} of the "soluble" PBC show a quiet broad static pattern similar to relaxing spectra; however, it is hard to conclude if the origin is an experimental broadening, folding shift or a true observation as in our case. Another nanoparticle size of the clusters may play a role as well.} \cite{Ito}. The results of $^{57}$Fe Mössbauer spectroscopy electrostatic monopole interaction yielded isomer shift values of $\delta=-0.146(10)$\,mms/s of Fe(II) and $\delta=0.408(30)$\,mm/s of Fe(III) at room temperature (see SI Fig. S10). These values of Fe(II) and Fe(III) are consistent with the value ranges of PBCs according to Samain \textit{et al.} \cite{Samain2013,Samain2} and Reguera \textit{et al.}\footnote{Note: Room temperature M\"ossbauer results by Reguera \textit{et al.} \cite{reguera_mossbauer_1992} are relative to sodium nitroprusside (Na$_2$[Fe(CN)$_5$NO]) which is  $\delta= -0.2576(14)$\,mm/s \cite{RUSANOV20091252} relative to $\alpha$-iron.}  \cite{reguera_mossbauer_1992}. In the following, we focus on the meaningful measurements at base temperature. At high temperature, charge transfer effects and diffusion or large polaron effects are discussed \cite{Ito}. 
Fig. \ref{figmoess} shows the two impactful $^{57}$Fe Mössbauer spectra at 4.2\,K and 2.5\,K. The electrostatic hyperfine parameter of hyperfine interactions and Fe species (Fe(II) $S=0$ green; Fe(III) $S=5/2$ red) are again fully consistent with AFCF \cite{Samain2013,ERICKSON19701195}, for details see below. The large line width of the ferric site is caused by the intrinsic local disorder of water molecules and vacancies that leads to a distribution of values of the quadrupole splitting of the \textit{Giese-salt} and in PBC, respectively \cite{greaves_site_2016}. At lower temperature, the Fe(III) species develops a quasi-static sextet down to 2.5\,K as often observed in molecular magnets and magnetic nanoparticles. Ferromagnetism and mixed-valence delocalization are discussed using a trapping-hopping model of lattice polarization \cite{DT9760001483} in PB. The shape of the red line is described by a Blume relaxation model \cite{chuev_multilevel_2012}. It originates from thermal activated quantum tunneling of the magnetic hyperfine field $B=51.99(51)$\,T at 2.5\,K. In this case, we used a Blume model with powder averaging and multi-level relaxation in contrast to single crystal studies of a two-level system \cite{braeuninger}. 
At base temperature, avoiding significant second order Doppler shift, the electric monopole interaction of Fe(III) is given by the isomer shift $\delta=0.53(1)$\,mm/s with respect to $\alpha$-Fe. 
The values obtained are consistent with Samain \textit{et al.} 2013 \cite{Samain2013} and Ito \textit{et al.} 1968 \cite{Ito}. 
The electric quadrupole interaction is represented by the largest component of the electric field gradient tensor $V_{zz}=7.1(2)$\,V/\AA$^2$ of Fe(III). The Blume model was modeled by a total moment quantum number $S=66$, a fluctuation rate $f=0.04(1)$\,MHz and a ratio of the single-macrospin energy barrier $KV$ and the thermal energy $k_BT$ of 
\begin{equation}
    \frac{KV}{k_BT}=4.08(2),
\end{equation}
where $K$ is the effective anisotropy energy density, $V$ the volume, $k_B$ is the Boltzmann constant, and $T$ the temperature. The model is based on a quantum mechanical description of a uniformly magnetized nanoparticle with the total spin $S$, and $2S + 1$ stochastic states of the projection of the spin onto the magnetic anisotropy axis $z$ with $m =-S, -S+1,..., S$, and the transitions between the states are caused by the transverse components of random fields \cite{chuev_multilevel_2012}.
Therefore, the relaxation pattern of the Blume model shows an effective magnetic short-range order including the confirmed ferromagnetism. AFCF is a coordination polymer comparable to a quasi-metal organic framework, at least from a Mössbauer perspective. Considering the ferromagnetic interaction of the spin centers, the effective cluster size corresponds to $66/2.5\approx26$ Fe(III) centers. A linear chain of double-exchanged Fe(III) by N$\equiv$C$-$Fe(II)$-$C$\equiv$N bridges  \cite{DT9760001483} yields an upper length of 27\,nm. Although  microscopy and LD examinations propose effective macroscopic particle sizes in the order of $\mu m$, the relaxing $^{57}$Fe Mössbauer spectra strongly confirm the result of \textit{D} $\approx$  19.1\,nm $\pm$ 2.3\,nm as estimated by XRD for crystal sizes. This shows especially that the agglomeration, the formation of larger particles by weak interactions of or loose bonds between smaller particles, is naturally not a long-range structurally ordered crystallization process. Therefore, \textit{Giese-salt} AFCF form magnetic and structural nano-like clusters with slow relaxing ferromagnetic short-range order at 2.5\,K. Our result is consistent with early neutron scattering results \cite{doi:10.1021/ic50206a032} and $^{57}$Fe Mössbauer measurement of the static regime \cite{Ito} in similar PBCs. An intriguing magnetic aspect explored is the potential magnetic-exchange interaction facilitated by the electronic coupling that occurs between cesium ions and the PB lattice structure. This interaction is proposed to result from the transfer of spin density via bonds directly to the cesium ions. This process highlights the possibility that cesium ions are not simply situated in interstitial positions but are instead electronically incorporated into the PB matrix, thereby becoming an integral part of the framework. The findings suggest a more complex role for cesium ions in the electronic landscape of the system \cite{koehler}. Magnetically, the additional counterion disorder resulting from K$^+$ and $\mathrm{NH}_4^+$ is expected to be negligible due to the ferromagnetic nature \cite{Ito}. Finally, it should be mentioned that an enlarged Fe(II)$-$C$\equiv$N$-$Fe(III) bond length may influence orbital overlap and ferromagnetic interaction and therefore the ferromagnetic cluster particle transition temperature and that the detailed ferromagnetic nature is still in controversy, e.g., proposed to be dominated by the coupling of nearest-neighbor high spin interaction \cite{middlemiss}.

\section*{Conclusions}\label{conclusion}
According to Giese \cite{gieserev}, the colloidal Prussian blue compound AFCF is a leading antidote to radiocesium in veterinary medicine, being 86-266 times more effective than the ion exchangers Bentonite or Bolus alba.
In modern spectroscopy, deep investigation, highlighted by using methods of nuclear spectroscopical chemistry, confirms the well-defined chemical properties. The term \textit{Giese-salt} appears to be used mainly in the European scientific literature, particularly in studies related to the remediation of nuclear accidents \cite{hoveeffect1990,mathiesen1990,Kaikkonen2000,Dresow}. 

A purified sample of the Prussian blue compound AFCF, synthesized in the same way as it is performed industrially for use as a radiocesium intoxication medication in veterinary medicine, was subjected to a detailed compositional and structural examination using various elemental and species analysis techniques. The purpose was to contribute to the goal of an improved understanding of the chemical and structural properties of veterinary medically applied AFCF, which is crucial for a better and deeper understanding of the biomolecular or, respectively, pharmaceutical mechanisms and effectiveness of PBC. Since PBCs always bear the challenge of distinct variability in structural details and all related characteristics, not only between PBCs of different composition but particularly also between compounds of basically the same composition but different origin (e.g. by means of synthesis processes), very often detailed individual examinations of specific PBCs of interests are required. This is the case of AFCF produced for veterinary medical application.

As essential for detailed compositional and structural examinations, the basic product identity of the AFCF sample as a "soluble" PB containing NH$_4^+$ was verified by rapid tests focusing on the macroscopic and chemical characteristics of the sample. Although in the additional light and electron microscopic as well as LD examinations particles in the whole micrometer magnitude were observed, the AFCF sample is a (sub)micrometer- to nanometer-sized powder as it is required for a "soluble" PB and the larger particles were just agglomerates of much smaller grains or structural crystals, as confirmed by results of XRD (structural clusters) and Mössbauer spectroscopic (magnetic clusters). In addition to this basic identity, the purity of the AFCF sample, especially with respect to the absence of the major by-product ammonium chloride (NH$_4$Cl), the successful dialysis purification procedure, was also crucial for further in-depth examinations to exclude significant interferences and matrix effects. This required purity was confirmed by the qualitative XRD results since all observed reflexes could be assigned to the basic crystal structure of PB\footnote{Nevertheless, the non-ideal $S$/$N$ ratio and, consequentially assumed, the limited LOD in the XRD examinations of this study (due to the applied XRD set-up, especially the for iron containing samples non-ideal Cu K$\mathrm{\alpha}$ radiation) must be considered. Thus, impurities of very small concentrations, negligible for veterinary use, cannot be excluded.}.
Furthermore, the ICP-OES analysis also revealed a sufficient purity grade of the sample. Although trace elements in the concentration magnitude of approximately 5200\,mg/kg with potassium as the main contributor (about 2200\,mg/kg), presumably originating from educt impurities and substituting, to a very small extent, NH$_4^+$ cations in the solid structure of the AFCF sample, were detected, greater effects on the significance of further in-depth compositional and structural examinations were not expected. Further investigations, in detail, the obtained Raman\footnote{Despite some smaller differences and variability which simply corresponded to various data and confusion in the literature.}, IR, and UV-vis spectroscopy results, as well as qualitative XRD phase analysis results structurally confirmed AFCF as a NH$_4^+$ containing PBC. In particular, the results of elemental analysis (ICP-OES and CHNS), XAFS and Mössbauer revealed much more variability and sample specific characteristics.

The ICP-OES and CHNS results demonstrated two aspects. First, the determined Fe, N, H, and C contents clearly suggested that AFCF is a hydrate and, thus, has a composition of the type NH$_4\{$Fe(III)[Fe(II)(CN)$_6]\}$ $\cdot$ x H$_2$O. This is in agreement with some reports in the literature which also describe "soluble" PBs as hydrates \cite{Samain2013, Riedel2007, Ware2008}. It should be emphasized that this information can be relevant for improved calculations of oral pharmaceutical doses in the veterinary medical application of PBC. However, for a complete consideration of this aspect, a precise knowledge of the amount of water x would be required. With regard to this topic, the results of this study were not finally clear. Depending on the determined element content, different potential values of x were calculated when comparing the specific element content with the value expected in the case of an ideal stoichiometric composition (for example, Fe: x $\approx$ 4, but H: x $\approx$ 1-2). Although it is known from the literature that an exact determination of the water content of PB is challenging \cite{Samain2013, MARTINEZALONSO2024}, attempts are planned to obtain more precise information by including thermogravimetric analysis (TGA) for future examinations of the purified AFCF sample. Secondly, which also limits the expressiveness of the obtained element concentration results for determining the water content of the sample, the determined N and C concentrations, more specifically their ratio strongly indicated differences from the ideal stoichiometry and structure in the form of the presence of defects in the structure of the AFCF sample. This is further confirmed by the ratios of the determined Fe and C as well as N contents and also by an effective large line width of the Fe$^{3+}$ HS site by M\"ossbauer spectroscopy, which represents a distribution of hyperfine parameters by local disorder. This would be in congruence with the studies mentioned in the introduction section (e.g. Kraft 2022 \cite{Kraft2022}) in which also "soluble" PBs are described as defect-containing structures.   

The XAFS measurements of the purified AFCF sample clearly reflected the structural characteristics of a PB in comparison to the measured spectra of the successfully synthesized and pre-characterized reference PB compounds PB I, II and III. However, two different major spectral features were found that seem to be specific to the AFCF sample. First, the slope steepness in the XANES area was right between those of the reference PBs. Although it was more similar to the steepness and shape of the slope of the "soluble" PB III, it cannot be taken as an additional significant proof that AFCF is also a "soluble" PB. This is just a further example of the variability described in the spectral characteristics of PBCs. It represents the AFCF samples Fe(II)$-$C$\equiv$N$-$Fe(III) system and sample specific deviations such as distortions, bond length differences, or vacancies of this ideal system \cite{Samain2013, Gervais2013}. Additionally, these slight slope steepness differences were even obtained for the two "insoluble" PB reference compounds PB I and II, which should have the same basic composition but were of different synthetic origin. Secondly, a significant edge shift towards lower energy as well as EXAFS oscillation shifts suggested significant enlarged bond lengths in the Fe(II)$-$C$\equiv$N$-$Fe(III) system of the AFCF sample. This aspect is of interest for an improved understanding of the pharmaceutical or biomolecular mechanisms of the veterinary medical applied AFCF because an enlarged Fe(II)$-$C$\equiv$N$-$Fe(III) system and, thus, enlarged octants can affect the ion exchange properties which are the essence of the mode of action of AFCF for radiocesium decorporation of animals. Additionally, this modified structural property may also include potential in biofunctionality. Hence, further investigation in the form of more detailed EXAFS analyses which enable the quantitative determination of specific bond distances should be performed in the examinations of the purified AFCF sample. 

As shown by low-temperature $^{57}$Fe Mössbauer spectroscopy or, more precisely, by multilevel relaxation on the rough Mössbauer time scale, the AFCF sample also exhibits ferromagnetic clusters with an upper length of about 27\,nm at 2.5\,K, here, the particle size can be defined microporous (1-2\,nm), mesoporous (2-50\,nm), or macroporous (50\,nm-50$\mu$m) \cite{Estelrich2021}. These cluster sizes are in relatively good accordance with the crystal sizes of \textit{D} $\approx$  19.1\,nm $\pm$ 2.3\,nm determined by the approximation of applying the Scherrer method to the XRD results. It is often discussed that the size of the cluster may influence the rate of radiocesium adsorption. In further examinations of the purified AFCF sample, instrumental broadening, which could not be determined within the scope of this study, must be taken into account and, additionally, an optimized XRD setup, for example, using Co K$\mathrm{\alpha}$ radiation that is better suited for iron-containing samples \cite{Mos2018}, should be employed to obtain improved results. This also counts for the XRD phase analysis results to get optimized $S$/$N$ ratios and LODs.

In total, the results of this study clearly demonstrate the challenging structural variability of PBCs, particularly reflected in spectral fine structures, and the imperative of individual and specific examinations of each PBC of interest in or for a specific application again. For example, this is important for Prussian blue nanoparticles, known for their biocompatibility and distinct characteristics, which have generated significant interest in biomedical research \cite{quin}, for example, in cancer therapy \cite{gao}, drug delivery \cite{zhang_2025}, imaging \cite{busquet}, and are still of important interest in radiocesium decontamination in case of future nuclear or radiological accidents.

\section*{Acknowledgements}
We would like to thank Willem F. Wolkers (\textit{Lower Saxony Centre for Biomedical Engineering, Implant Research and Development} (\textit{NIFE}), University of Veterinary Medicine Hannover Foundation) for providing the ATR-IR spectrometer and supporting the measurements of this study. We thank Beatriz Del Rocio Dörrie Delgado (Institute of Sanitary Engineering and Waste Management, Leibniz University Hannover) and Michael Stern (Institute of Physiology and Cell Biology, University of Veterinary Medicine Hannover Foundation) for great support with the light microscopy examinations. Furthermore, we also want to thank \textit{Retsch GmbH} for the LD-based particle size distribution measurements. Finally, special thanks go to W. W. Giese ($\dagger$ 2023) for historic original equipment.  We acknowledge financial support by the Open Access Publication Fund of the
University of Veterinary Medicine Hannover, Foundation.

\section*{Conflicts of Interest}

There are no conflicts to declare.

\section*{Supplementary Information}
The PDF document offers a detailed description that presents the methodologies and processes utilized in the synthesis, along with initial characterizations of the reference compounds PB I, II, and III. Moreover, it contains a comprehensive compilation of results derived from the associated AFCF analytical procedures, providing in-depth insights into the study's findings.



\bibliography{bibliography}

@article{GIESE1988363,
title = {Ammonium-ferric-cyano-ferrate(II) (AFCF) as an effective antidote against radiocaesium burdens in domestic animals and animal derived foods},
journal = {British Veterinary Journal},
volume = {144},
number = {4},
pages = {363-369},
year = {1988},
issn = {0007-1935},
doi = {https://doi.org/10.1016/0007-1935(88)90065-6},
url = {https://www.sciencedirect.com/science/article/pii/0007193588900656},
author = {W.W. Giese}
}

@article{reguera_mossbauer_1992,
	title = {Mössbauer spectroscopic study of {Prussian} {Blue} from different provenances},
	volume = {73},
	issn = {0304-3834, 1572-9540},
	url = {http://link.springer.com/10.1007/BF02418604},
	doi = {10.1007/BF02418604},
	language = {en},
	number = {3-4},
	journal = {Hyperfine Interactions},
	author = {Reguera, E. and Fernández-Bertrán, J. and Dago, A. and Díaz, C.},
	month = oct,
	year = {1992},
	pages = {295--308},
}

@article{doi:10.1021/ac60214a047,
author = {Savitzky, Abraham. and Golay, M. J. E.},
title = {Smoothing and Differentiation of Data by Simplified Least Squares Procedures.},
journal = {Analytical Chemistry},
volume = {36},
number = {8},
pages = {1627-1639},
year = {1964},
doi = {10.1021/ac60214a047}

}

@article{Moretti2018RamanSO,
  title={Raman spectroscopy of the photosensitive pigment Prussian blue},
  author={Giuliano Moretti and Claire Gervais},
  journal={Journal of Raman Spectroscopy},
  year={2018},
  volume={49},
  pages={1198-1204},
doi = {https://doi.org/10.1002/jrs.5366}
}

@article{Monthiers,
  title={Des cyanures doubles},
  author={Monthiers, J. H..},
  journal={These de pharmacie de Paris},
  year={1847},
  volume={},
  pages={},
  url={}
}

@book{pigment,
author = {Eastaugh, Nicholas and Walsh, V and Chaplin, Tracey and Siddall, Ruth},
year = {1974},
month = {01},
pages = {},
title = {Pigment Compendium: A Dictionary and Optical Microscopy of Historic Pigments}
}

@book{riffault_des_hetres_practical_1874,
	location = {Philadelphia},
	title = {A practical treatise on the manufacture of colors for painting. Comprising the origin, definition, and classification of colors; the treatment of the raw materials etc},
	publisher = {H.C. Baird Philadelphia},
	author = {Riffault des Hêtres, Jean René Denis and Vergnaud, A. D. and Toussaint, G. Alvar and Malepeyre, F. and Fesquet, A. A.},
	date = {1874},
	note = {Section: page 659 diagrams 24 cm},
}

@article{hanawalt,
author = {Hanawalt, J. D. and Rinn, H. W. and Frevel, L. K.},
title = {Chemical Analysis by X-Ray Diffraction},
journal = {Industrial \& Engineering Chemistry Analytical Edition},
volume = {10},
number = {9},
pages = {457-512},
year = {1938},
doi = {10.1021/ac50125a001}

}

@article{safetygiesesatl,
author = {{ EFSA Panel on Additives and Products or Substances used in Animal Feed (FEEDAP) } and Bampidis, Vasileios and Azimonti, Giovanna and Bastos, Maria de Lourdes and Christensen, Henrik and Dusemund, Birgit and Fašmon Durjava, Mojca and Kouba, Maryline and López-Alonso, Marta and López Puente, Secundino and Marcon, Francesca and Mayo, Baltasar and Pechová, Alena and Petkova, Mariana and Ramos, Fernando and Sanz, Yolanda and Villa, Roberto Edoardo and Woutersen, Ruud and Innocenti, Matteo Lorenzo and Pizzo, Fabiola and Galobat, Jaume and Holczknecht, Orsolya and Bories, Georges and Gropp, Jürgen and Nebbia, Carlo and Aquilina, Gabriele},
title = {Safety and efficacy of a feed additive consisting of ferric (III) ammonium hexacyanoferrate (II) for ruminants (domestic and wild), calves prior the start of rumination, lambs prior the start of rumination, kids prior the start of rumination and pigs (domestic and wild) (Honeywell Specialty Chemicals Seelze GmbH)},
journal = {EFSA Journal},
volume = {19},
number = {6},
pages = {e06628},
doi = {https://doi.org/10.2903/j.efsa.2021.6628},
year = {2021}
}

@article{m_giese_1988, 
title={Synthesis, effectiveness and metabolic fate in cows of the caesium complexing compound ammonium ferric hexacyanoferrate labelled with 14C}, volume={55}, DOI={10.1017/S0022029900025796}, number={1}, journal={Journal of Dairy Research}, publisher={Cambridge University Press}, author={Arnaud, Maurice J. and Clement, Charles and Getaz, Françoise and Tannhauser, Fritz and Schoenegge, Rainer and Blum, Jürg and Giese, Werner}, year={1988}, pages={1–13}
}

@article{DABURON199173,
title = {Radiocaesium transfer to Ewes Fed contaminated hay after the chernobyl accident: Effect of vermiculite and AFCF (Ammonium Ferricyanoferrate) as countermeasures},
journal = {Journal of Environmental Radioactivity},
volume = {14},
number = {1},
pages = {73-84},
year = {1991},
issn = {0265-931X},
doi = {https://doi.org/10.1016/0265-931X(91)90016-9},
url = {https://www.sciencedirect.com/science/article/pii/0265931X91900169},
author = {F. Daburon and Y. Archimbaud and J. Cousi and G. Fayart and D. Hoffschir and I. Chevallereau and H. {Le Creff} and L. Gueguen}
}

@article{ramanlib,
author = {Fremout, Wim and Saverwyns, Steven},
title = {Identification of synthetic organic pigments: the role of a comprehensive digital Raman spectral library},
journal = {Journal of Raman Spectroscopy},
volume = {43},
number = {11},
pages = {1536-1544},
doi = {https://doi.org/10.1002/jrs.4054},
year = {2012}
}

@article{Kraft2022,
author = {Kraft, Alexander},
title = {Berliner Blau im 21. Jahrhundert},
journal = {Chemie in unserer Zeit},
volume = {56},
number = {2},
pages = {110-123},
doi = {https://doi.org/10.1002/ciuz.202100032},
year = {2022}
}

@Article{Buser1972,
author ="Buser, H. J. and Ludi, A. and Petter, W. and Schwarzenbach, D.",
title  ="Single-crystal study of Prussian Blue: Fe4[Fe(CN)6]2{,} 14H2O",
journal  ="J. Chem. Soc.{,} Chem. Commun.",
year  ="1972",
issue  ="23",
pages  ="1299-1299",
publisher  ="The Royal Society of Chemistry",
doi  ="10.1039/C39720001299"
}

@article{Buser1977,
author = {Buser, H. J. and Schwarzenbach, D. and Petter, W. and Ludi, A.},
title = {The crystal structure of Prussian Blue:  Fe4[Fe(CN)6]3.xH2O},
journal = {Inorganic Chemistry},
volume = {16},
number = {11},
pages = {2704-2710},
year = {1977},
doi = {10.1021/ic50177a008}
    
    

}

@article{Sharma2014,
    author = {Sharma, V. K. and Mitra, S. and Thakur, N. and Yusuf, S. M. and Juranyi, Fanni and Mukhopadhyay, R.},
    title = "{Dynamics of water in prussian blue analogues: Neutron scattering study}",
    journal = {Journal of Applied Physics},
    volume = {116},
    number = {3},
    pages = {034909},
    year = {2014},
    month = {07},
    issn = {0021-8979},
    doi = {10.1063/1.4890722},
}

@Article{Estelrich2021,
AUTHOR = {Estelrich, Joan and Busquets, Maria Antònia},
TITLE = {Prussian Blue: A Safe Pigment with Zeolitic-Like Activity},
JOURNAL = {International Journal of Molecular Sciences},
VOLUME = {22},
YEAR = {2021},
NUMBER = {2},
ARTICLE-NUMBER = {780},
URL = {https://www.mdpi.com/1422-0067/22/2/780},
PubMedID = {33467391},
ISSN = {1422-0067},
DOI = {10.3390/ijms22020780}
}

@Article{Gervais2013,
author ="Gervais, Claire and Languille, Marie-Angélique and Réguer, Solenn and Gillet, Martine and Pelletier, Sébastien and Garnier, Chantal and Vicenzi, Edward P. and Bertrand, Loïc",
title  ="Why does Prussian blue fade? Understanding the role(s) of the substrate",
journal  ="J. Anal. At. Spectrom.",
year  ="2013",
volume  ="28",
issue  ="10",
pages  ="1600-1609",
publisher  ="The Royal Society of Chemistry",
doi  ="10.1039/C3JA50025J",
url  ="http://dx.doi.org/10.1039/C3JA50025J"
}

@article{Samain2013,
author = {Samain, Louise and Grandjean, Fernande and Long, Gary J. and Martinetto, Pauline and Bordet, Pierre and Strivay, David},
title = {Relationship between the Synthesis of Prussian Blue Pigments, Their Color, Physical Properties, and Their Behavior in Paint Layers},
journal = {The Journal of Physical Chemistry C},
volume = {117},
number = {19},
pages = {9693-9712},
year = {2013},
doi = {10.1021/jp3111327}
    

}

@Article{Samain2,
author ="Samain, Louise and Silversmit, Geert and Sanyova, Jana and Vekemans, Bart and Salomon, Hélène and Gilbert, Bernard and Grandjean, Fernande and Long, Gary J. and Hermann, Raphaël P. and Vincze, Laszlo and Strivay, David",
title  ="Fading of modern Prussian blue pigments in linseed oil medium",
journal  ="J. Anal. At. Spectrom.",
year  ="2011",
volume  ="26",
issue  ="5",
pages  ="930-941",
publisher  ="The Royal Society of Chemistry",
doi  ="10.1039/C0JA00234H",
url  ="http://dx.doi.org/10.1039/C0JA00234H"}

@article{Bueno2008,
author = {Bueno, Paulo Roberto and Ferreira, Fabio Furlan and Giménez-Romero, David and Oliveira Setti, Grazielle and Faria, Ronaldo Censi and Gabrielli, Claude and Perrot, Hubert and Garcia-Jareño, José Juan and Vicente, Francisco},
title = {Synchrotron Structural Characterization of Electrochemically Synthesized Hexacyanoferrates Containing K+: A Revisited Analysis of Electrochemical Redox},
journal = {The Journal of Physical Chemistry C},
volume = {112},
number = {34},
pages = {13264-13271},
year = {2008},
doi = {10.1021/jp802070f},
    

}

@article{Kjeldgaard2021,
author = {Kjeldgaard, Solveig  and Dugulan, Iulian  and Mamakhel, Aref  and Wagemaker, Marnix  and Iversen, Bo Brummerstedt  and Bentien, Anders },
title = {Strategies for synthesis of Prussian blue analogues},
journal = {Royal Society Open Science},
volume = {8},
number = {1},
pages = {201779},
year = {2021},
doi = {10.1098/rsos.201779}
}

@Article{Mamontova2022,
author ="Mamontova, Ekaterina and Salles, Fabrice and Guari, Yannick and Larionova, Joulia and Long, Jérôme",
title  ="Post-synthetic modification of Prussian blue type nanoparticles: tailoring the chemical and physical properties",
journal  ="Inorg. Chem. Front.",
year  ="2022",
volume  ="9",
issue  ="15",
pages  ="3943-3971",
publisher  ="The Royal Society of Chemistry",
doi  ="10.1039/D2QI01068B",
url  ="http://dx.doi.org/10.1039/D2QI01068B"
}

@article{Curdt2022,
author={Curdt, Franziska
and Haase, Katrin
and Ziegenbalg, Laura
and Greb, Helena
and Heyers, Dominik
and Winklhofer, Michael},
title={Prussian blue technique is prone to yield false negative results in magnetoreception research},
journal={Scientific Reports},
year={2022},
month={May},
day={25},
volume={12},
number={1},
pages={8803},
issn={2045-2322},
doi={10.1038/s41598-022-12398-9},
url={https://doi.org/10.1038/s41598-022-12398-9}
}

@article{Ludi1981,
author = {Ludi, Andreas},
title = {Prussian blue, an inorganic evergreen},
journal = {Journal of Chemical Education},
volume = {58},
number = {12},
pages = {1013},
year = {1981},
doi = {10.1021/ed058p1013}

}

@article{Ware2008,
author = {Ware, Mike},
title = {Prussian Blue: Artists' Pigment and Chemists' Sponge},
journal = {Journal of Chemical Education},
volume = {85},
number = {5},
pages = {612},
year = {2008},
doi = {10.1021/ed085p612}
    
    

}

@article{Kaikkonen2000,
author ="Kaikkonen, Matti and Lehto, Jukka",
title  ="Coprecipitating ammonium–iron(iii)–hexacyanoferrate(ii) from aqueous dispersion with albumin and trichloroacetic acid",
journal  ="Analyst",
year  ="2000",
volume  ="125",
issue  ="5",
pages  ="855-859",
publisher  ="The Royal Society of Chemistry",
doi  ="10.1039/B001154L",
}

@book{Gade1998_Kap2,
author = {Gade, Lutz H.},
publisher = {John Wiley and Sons, Ltd},
isbn = {9783527663927},
title = {Koordinationschemie},
chapter = {2},
pages = {3-24},
address = {Germany},
doi = {https://doi.org/10.1002/9783527663927.ch2},
year = {1998}
}

@book{Beyer2012Kap1,
author="Beyer, Lothar
and Cornejo, Jorge Angulo",
title="Koordinationschemie: Grundlagen-Synthesen-Anwendungen",
year="2012",
publisher="Vieweg+Teubner Verlag",
address="Wiesbaden",
pages="1--116",
chapter = "1",
isbn="978-3-8348-8343-8",
doi="10.1007/978-3-8348-8343-8_1",
}

@inbook{Janiak+2007+381+580,
title = {3 Komplex-/ Koordinationschemie},
booktitle = {Moderne Anorganische Chemie},
author = {Christoph Janiak},
editor = {Erwin Riedel},
publisher = {De Gruyter},
address = {Berlin, Boston},
pages = {381--580},
doi = {doi:10.1515/9783110206852-004},
isbn = {9783110206852},
year = {2007},
lastchecked = {2023-11-26}
}

@book{Wiberg+2008+1635+1680,
title = {Lehrbuch der Anorganischen Chemie},
author = {Nils Wiberg},
editor = {Arnold F. Holleman},
publisher = {De Gruyter},
address = {Berlin, Boston},
pages = {1635--1680},
chapter = {XXIX},
doi = {doi:10.1515/9783110206845-031},
isbn = {9783110206845},
year = {2008},
lastchecked = {2023-11-26}
}

@book{Riedel2007,
title = {Anorganische Chemie},
author = {Erwin Riedel and Christoph Janiak},
publisher = {De Gruyter},
address = {Berlin, Boston},
pages = {659--886},
doi = {doi:10.1515/9783110206869-006},
isbn = {9783110206869},
year = {2007},
chapter = {5},
lastchecked = {2023-11-26}
}

@book{Latscha2004,
author="Latscha, Hans Peter
and Linti, Gerald Walter
and Klein, Helmut Alfons",
title="Analytische Chemie: Chemie---Basiswissen III",
year="2004",
publisher="Springer Berlin Heidelberg",
address="Berlin, Heidelberg",
pages="9--114",
isbn="978-3-642-18493-2",
doi="10.1007/978-3-642-18493-2_2",
}

@Inbook{Vandenberghe2013,
author="Vandenberghe, Robert E.
and De Grave, Eddy",
editor="Yoshida, Yutaka
and Langouche, Guido",
title="Application of M{\"o}ssbauer Spectroscopy in Earth Sciences",
bookTitle="M{\"o}ssbauer Spectroscopy: Tutorial Book",
year="2013",
publisher="Springer Berlin Heidelberg",
address="Berlin, Heidelberg",
pages="91--185",
isbn="978-3-642-32220-4",
doi="10.1007/978-3-642-32220-4_3",
}

@Inbook{Gütlich2013,
author="G{\"u}tlich, Philipp
and Garcia, Yann",
editor="Yoshida, Yutaka
and Langouche, Guido",
title="Chemical Applications of M{\"o}ssbauer Spectroscopy",
bookTitle="M{\"o}ssbauer Spectroscopy: Tutorial Book",
year="2013",
publisher="Springer Berlin Heidelberg",
address="Berlin, Heidelberg",
pages="23--89",
isbn="978-3-642-32220-4",
doi="10.1007/978-3-642-32220-4_2",
}

@article{Nobrega1996,
title = {Flow-injection spectrophotometric determination of ascorbic acid in pharmaceutical products with the Prussian Blue reaction},
journal = {Talanta},
volume = {43},
number = {6},
pages = {971-976},
year = {1996},
issn = {0039-9140},
doi = {https://doi.org/10.1016/0039-9140(95)01830-1},
url = {https://www.sciencedirect.com/science/article/pii/0039914095018301},
author = {Joaquim A. Nóbrega and Gisele S. Lopes}
}

@article{Manabe2020,
title = {Stabilization of Prussian blue using copper sulfate for eliminating radioactive cesium from a high pH solution and seawater},
journal = {Journal of Hazardous Materials},
volume = {386},
pages = {121979},
year = {2020},
issn = {0304-3894},
doi = {https://doi.org/10.1016/j.jhazmat.2019.121979},
url = {https://www.sciencedirect.com/science/article/pii/S0304389419319338},
author = {Shoichi Manabe and Vipin {Adavan Kiliyankil} and Tsuguo Kumashiro and Shoichi Takiguchi and Bunshi Fugetsu and Ichiro Sakata}
}

@article{Adhikamsetty2009,
author = {Adhikamsetty, R. K. and Jonnalagadda, S. B.},
title = {Kinetics and mechanism of Prussian blue formation},
journal = {Bulletin of the Chemical Society of Ethiopia},
volume = {23},
number = {1},
pages = {47-54},
year = {2009},
doi = {10.4314/bcse.v23i1.21297}
}

@article{Roth2022,
author = {Roth, Klaus},
title = {Berliner Blau – Entdecker und Verräter},
journal = {Chemie in unserer Zeit},
volume = {56},
number = {1},
pages = {34-49},
doi = {https://doi.org/10.1002/ciuz.202100033},
year = {2022}
}

@article{Gournis2010,
title = {A two-dimensional magnetic hybrid material based on intercalation of a cationic Prussian blue analog in montmorillonite nanoclay},
journal = {Journal of Colloid and Interface Science},
volume = {348},
number = {2},
pages = {393-401},
year = {2010},
issn = {0021-9797},
doi = {https://doi.org/10.1016/j.jcis.2010.04.068},
url = {https://www.sciencedirect.com/science/article/pii/S0021979710004698},
author = {Dimitrios Gournis and Christina Papachristodoulou and Enrico Maccallini and Petra Rudolf and Michael A. Karakassides and Dimitrios T. Karamanis and Marie-Hélène Sage and Thomas T.M. Palstra and Jean-François Colomer and Konstantinos D. Papavasileiou and Vasilios S. Melissas and Nicolaos H. Gangas}
}

@inbook{Robine1906,
author = {Robine, R. and Lenglen, M.},
title = {The Manufacture of Prussian Blue and various other compounds},
booktitle = {The Cyanide Industry - Theoretically and practically considered},
year = {1906}, 
publisher = {John Wiley Sons},
address = {New York},
pages = {288--293}
}

@article{Welgama2023,
author = {Welgama, Heshali K. and Crawley, Matthew R. and McKone, James R. and Cook, Timothy R.},
title = {Investigations of Nanoparticle Suspensions of Prussian Blue and Its Copper Analogue: Amine Functionalization and Electrochemical Studies},
journal = {Inorganic Chemistry},
volume = {62},
number = {4},
pages = {1455-1465},
year = {2023},
doi = {10.1021/acs.inorgchem.2c03545}
    
    

}

@Article{Carniato2020,
AUTHOR = {Carniato, Fabio and Gatti, Giorgio and Vittoni, Chiara and Katsev, Andrey M. and Guidotti, Matteo and Evangelisti, Claudio and Bisio, Chiara},
TITLE = {More Efficient Prussian Blue Nanoparticles for an Improved Caesium Decontamination from Aqueous Solutions and Biological Fluids},
JOURNAL = {Molecules},
VOLUME = {25},
YEAR = {2020},
NUMBER = {15},
ARTICLE-NUMBER = {3447},
URL = {https://www.mdpi.com/1420-3049/25/15/3447},
PubMedID = {32751159},
ISSN = {1420-3049},
DOI = {10.3390/molecules25153447}
}

@article{doi:10.1021/ic50206a032,
author = {Herren, F. and Fischer, P. and Ludi, A. and Haelg, W.},
title = {Neutron diffraction study of Prussian Blue, Fe4[Fe(CN)6]3.xH2O. Location of water molecules and long-range magnetic order},
journal = {Inorganic Chemistry},
volume = {19},
number = {4},
pages = {956-959},
year = {1980},
doi = {10.1021/ic50206a032}
    
    

}

@article{Momma:db5098,
author = "Momma, Koichi and Izumi, Fujio",
title = "{{\it VESTA3} for three-dimensional visualization of crystal, volumetric and morphology data}",
journal = "Journal of Applied Crystallography",
year = "2011",
volume = "44",
number = "6",
pages = "1272--1276",
month = "Dec",
doi = {10.1107/S0021889811038970},
url = {https://doi.org/10.1107/S0021889811038970},
}

@article{doi:10.4491/eer.2018.177,
author = {Wi Hyobin,Kang Sung-Won,Hwang Yuhoon},
title = {Immobilization of Prussian blue nanoparticles in acrylic acid-surface functionalized poly(vinyl alcohol) sponges for cesium adsorption},
journal = {Environmental Engineering Research},
volume = {24},
number = {1},
pages = {173-179},
year = {2019},
doi = {10.4491/eer.2018.177},
}

@Article{Deng2023,
AUTHOR = {Deng, Tong and Garg, Vivek and Bradley, Michael S. A.},
TITLE = {Electrostatic Charging of Fine Powders and Assessment of Charge Polarity Using an Inductive Charge Sensor},
JOURNAL = {Nanomanufacturing},
VOLUME = {3},
YEAR = {2023},
NUMBER = {3},
PAGES = {281--292},
URL = {https://www.mdpi.com/2673-687X/3/3/18},
ISSN = {2673-687X},
DOI = {10.3390/nanomanufacturing3030018}
}

@article{MARTINEZALONSO2024,
title = {Physicochemical and pharmacotechnical characterization of Prussian blue for future Prussian blue oral dosage forms formulation},
journal = {Heliyon},
volume = {10},
number = {2},
pages = {e24284},
year = {2024},
issn = {2405-8440},
doi = {https://doi.org/10.1016/j.heliyon.2024.e24284},
url = {https://www.sciencedirect.com/science/article/pii/S2405844024003153},
author = {Borja Martínez-Alonso and Norma S. {Torres Pabón} and María Isabel Fernández-Bachiller and Guillermo Torrado Durán and Rocío González Crespo and Carlos F. Torrado-Salmerón and Antonio Juberías Sánchez and M. Ángeles {Peña Fernández}}
}

@article{Mos2018,
author = {Yvonne M. Mos, Arnold C. Vermeulen, Cees J. N. Buisman and Jan Weijma},
title = {X-Ray Diffraction of Iron Containing Samples: The Importance of a Suitable Configuration},
journal = {Geomicrobiology Journal},
volume = {35},
number = {6},
pages = {511--517},
year = {2018},
publisher = {Taylor \& Francis},
doi = {10.1080/01490451.2017.1401183},
    

}

@book{spiess2009moderne,
  title={Moderne R{\"o}ntgenbeugung: R{\"o}ntgendiffraktometrie f{\"u}r Materialwissenschaftler, Physiker und Chemiker},
  author={Spiess, L. and Teichert, G. and Schwarzer, R. and Behnken, H. and Genzel, C.},
  isbn={9783835101661},
  series={Vieweg Studium},
  year={2009},
  publisher={Vieweg+Teubner Verlag},
  address = {Wiesbaden}
}

@article{Agrisuelas_2009,
doi = {10.1149/1.3177258},
url = {https://dx.doi.org/10.1149/1.3177258},
year = {2009},
month = {jul},
publisher = {The Electrochemical Society, Inc.},
volume = {156},
number = {10},
pages = {P149},
author = {Jeronimo Agrisuelas and Jose Juan García-Jareño and David Gimenez-Romero and Francisco Vicente},
title = {Insights on the Mechanism of Insoluble-to-Soluble Prussian Blue Transformation},
journal = {Journal of The Electrochemical Society}
}

@article{Monteiro2017,
author = {Monteiro, Marcio C. and Toledo, Kalil C. F. and Pires, Bruno M. and Wick, René and Bonacin, Juliano A.},
title = {Improvement in Efficiency of the Electrocatalytic Reduction of Hydrogen Peroxide by Prussian Blue Produced from the [Fe(CN)5(mpz)]2– Complex},
journal = {European Journal of Inorganic Chemistry},
volume = {2017},
number = {13},
pages = {1979-1988},
doi = {https://doi.org/10.1002/ejic.201601540},
year = {2017}
}

@Article{Onoe2021,
author ="Onoe, Jun and Watanabe, Shinta and Masuda, Hideki and Inaba, Yusuke and Harigai, Miki and Takeshita, Kenji",
title  ="The uptake mechanism of palladium ions into Prussian-blue nanoparticles in a nitric acid solution toward application for the recycling of precious metals from electronic and nuclear wastes",
journal  ="RSC Adv.",
year  ="2021",
volume  ="11",
issue  ="34",
pages  ="20701-20707",
publisher  ="The Royal Society of Chemistry",
doi  ="10.1039/D1RA01794B"
}

@article{Watanabe2016,
    author = {Watanabe, Shinta and Sawada, Yuki and Nakaya, Masato and Yoshino, Masahito and Nagasaki, Takanori and Kameyama, Tatsuya and Torimoto, Tsukasa and Inaba, Yusuke and Takahashi, Hideharu and Takeshita, Kenji and Onoe, Jun},
    title = "{Intra- and inter-atomic optical transitions of Fe, Co, and Ni ferrocyanides studied using first-principles many-electron calculations}",
    journal = {Journal of Applied Physics},
    volume = {119},
    number = {23},
    pages = {235102},
    year = {2016},
    month = {06},
    issn = {0021-8979},
    doi = {10.1063/1.4954070},
}

@article{YANG2008,
title = {Quantitative determination of thallium binding to ferric hexacyanoferrate: Prussian blue},
journal = {International Journal of Pharmaceutics},
volume = {353},
number = {1},
pages = {187-194},
year = {2008},
issn = {0378-5173},
doi = {https://doi.org/10.1016/j.ijpharm.2007.11.031},
url = {https://www.sciencedirect.com/science/article/pii/S0378517307009672},
author = {Yongsheng Yang and Patrick J. Faustino and Joseph J. Progar and Charles R. Brownell and Nakissa Sadrieh and Joan C. May and Eldon Leutzinger and David A. Place and Eric P. Duffy and Lawrence X. Yu and Mansoor A. Khan and Robbe C. Lyon}
}

@article{Altagracia-Martínez2012,
author = {Altagracia-Martinez, M. and Kravzov-Jinich, J. and Martínez-
Núñez, J. M. and Ríos-Castañeda, C. and López-Naranjo, F.},
title = "{Prussian blue as an antidote for radioactive thallium and cesium poisoning}",
journal = {Orphan Drugs: Research and Reviews},
volume = {119},
number = {2},
pages = {13-21},
year = {2012},
doi = {10.2147/ODRR.S31881}




}

@article{Melvin1962,
author = {Robin, Melvin B.},
title = {The Color and Electronic Configurations of Prussian Blue},
journal = {Inorganic Chemistry},
volume = {1},
number = {2},
pages = {337-342},
year = {1962},
doi = {10.1021/ic50002a028}

}

@article{RUSANOV20091252,
title = {Determination of the Mössbauer parameters of rare-earth nitroprussides: Evidence for new light-induced magnetic excited state (LIMES) in nitroprussides},
journal = {Journal of Solid State Chemistry},
volume = {182},
number = {5},
pages = {1252-1259},
year = {2009},
issn = {0022-4596},
doi = {https://doi.org/10.1016/j.jssc.2009.02.023},
url = {https://www.sciencedirect.com/science/article/pii/S0022459609000814},
author = {V. Rusanov and S. Stankov and A. Ahmedova and A.X. Trautwein}
}

@article{NAWAR2023,
title = {Impedance spectroscopy and conduction mechanism analysis of bulk nanostructure Prussian blue pellets},
journal = {Materials Chemistry and Physics},
volume = {306},
pages = {128000},
year = {2023},
issn = {0254-0584},
doi = {https://doi.org/10.1016/j.matchemphys.2023.128000},
url = {https://www.sciencedirect.com/science/article/pii/S0254058423007083},
author = {Ahmed M. Nawar and Ahmed A. Alzharani}
}

@article{Penche2022,
title = {Transition Metal Hexacyanoferrate(II) Complexes as Catalysts in the Ring-Opening Copolymerization of CO2 and Propylene Oxide},
journal = {Topics in Catalysis},
volume = {65},
pages = {1541-1555},
year = {2022},
doi = {10.1007/s11244-022-01628-z},
url = {https://doi.org/10.1007/s11244-022-01628-z},
author = {Penche, Guillermo and González-Marcos, M. Pilar and González-Velasco, Juan R.}
}

@article{Vahur2016,
title = {ATR-FT-IR spectral collection of conservation materials in the extended region of 4000-80 cm-1},
journal = {Analytical and Bioanalytical Chemistry},
volume = {408},
pages = {3373-3379},
year = {2016},
doi = {10.1007/s00216-016-9411-5},
url = {https://doi.org/10.1007/s00216-016-9411-5},
author = {Vahur, Signe and Teearu, Anu and Peets, Pilleriin and Joosu, Lauri and Leito, Ivo}
}

@article{Miller1952,
author = {Miller, F. A. and Wilkins, C. H.},
title = {Infrared Spectra and Characteristic Frequencies of Inorganic Ions},
journal = {Analytical Chemistry},
volume = {24},
number = {8},
pages = {1253-1294},
year = {1952},
doi = {10.1021/ac60068a007},

URL = { 
    
        https://doi.org/10.1021/ac60068a007
    
    

},
eprint = { 
    
        https://doi.org/10.1021/ac60068a007
    
    

}

}

@book{Hesse2002,
  title={Spektroskopische Methoden in der organischen Chemie},
  author={Hesse, Manfred and Meier, Herbert and Zeeh, Bernd},
  isbn={3135761061},
  year={2002},
  publisher={Georg Thieme Verlag},
  address={Stuttgart}
}

@book{Bethge2008,
  title={Kernphysik},
  author={Bethge, Klaus and Walter, Gertrud and Wiedemann, Bernhard},
  isbn={9783540745662},
  year={2008},
  publisher={Springer Verlag},
  address={Berlin Heidelberg}
}

@book{Bunker_2010, 
place={Cambridge}, 
title={Introduction to XAFS: A Practical Guide to X-ray Absorption Fine Structure Spectroscopy}, 
publisher={Cambridge University Press}, 
author={Bunker, Grant},
year={2010},
address={Cambridge}}

@Inbook{Schnohr2015,
author="Schnohr, Claudia  S.
and Ridgway, Mark  C.",
editor="Schnohr, Claudia S.
and Ridgway, Mark C.",
title="Introduction to X-Ray Absorption Spectroscopy",
bookTitle="X-Ray Absorption Spectroscopy of Semiconductors",
year="2015",
publisher="Springer Berlin Heidelberg",
address="Berlin, Heidelberg",
pages="1--26",
isbn="978-3-662-44362-0",
doi="10.1007/978-3-662-44362-0_1"
}

@inbook{Kas2016,
author = {Kas, Joshua J. and Jorissen, Kevin and Rehr, John J.},
publisher = {John Wiley and Sons, Ltd},
isbn = {9781118844243},
title = {Real-Space Multiple-Scattering Theory of X-Ray Spectra},
booktitle = {X‐Ray Absorption and X‐Ray Emission Spectroscopy},
chapter = {3},
pages = {51-72},
doi = {https://doi.org/10.1002/9781118844243.ch3},
year = {2016},
address = {Chichester, West Sussex}
}

@inbook{Joly2016,
author = {Joly, Yves and Grenier, Stéphane},
publisher = {John Wiley and Sons, Ltd},
isbn = {9781118844243},
title = {Theory of X-Ray Absorption Near Edge Structure},
booktitle = {X‐Ray Absorption and X‐Ray Emission Spectroscopy},
chapter = {4},
pages = {73-97},
doi = {https://doi.org/10.1002/9781118844243.ch4},
year = {2016},
address = {Chichester, West Sussex}
}

@article{Wilke2001,
url = {https://doi.org/10.2138/am-2001-5-612},
title = {Oxidation state and coordination of Fe in minerals: An Fe K-XANES spectroscopic study},
title = {},
author = {Max Wilke and François Farges and Pierre-Emmanuel Petit and Gordon E. Brown and François Martin},
pages = {714--730},
volume = {86},
number = {5-6},
journal = {American Mineralogist},
doi = {doi:10.2138/am-2001-5-612},
year = {2001},
lastchecked = {2024-05-14}
}

@article{Matsukawa_1978,
doi = {10.7567/JJAPS.17S2.184},
url = {https://dx.doi.org/10.7567/JJAPS.17S2.184},
year = {1978},
month = {jan},
publisher = {},
volume = {17},
number = {S2},
pages = {184},
author = {Tokuo Matsukawa and Masayoshi Obashi and Shun-ichi Nakai and Chikara Suoiura},
title = {The K-Absorption Spectra of FeS2, CoS2 and NiS2},
journal = {Japanese Journal of Applied Physics}
}

@article{PETIAU1988237,
title = {K X-ray absorption spectra and electronic structure of chalcopyrite CuFeS2},
journal = {Materials Science and Engineering: B},
volume = {1},
number = {3},
pages = {237-249},
year = {1988},
issn = {0921-5107},
doi = {https://doi.org/10.1016/0921-5107(88)90004-9},
url = {https://www.sciencedirect.com/science/article/pii/0921510788900049},
author = {J. Petiau and Ph. Sainctavit and G. Calas}
}

@article{Lennie1996,
title = {Spectroscopic studies on iron sulfide formation and phase relations at low temperatures},
journal = {Geochemical Society Special Publication},
volume = {5},
pages = {117-132},
year = {1996},
author = {Lennie, A. R. and Vaughan, D. J.}
}

@article{Yamashige2005,
title = {Electronic structure analysis of iron(III)-porphyrin complexes by X-ray absorption spectra at the C, N and Fe K-edges},
journal = {Analytical Sciences},
volume = {21},
pages = {309-314},
year = {2005},
author = {Yamashige, Hisao and Matsuo, Shuji and Kurisaki, Tsutomu and Perera, Rupert C. C. and Wakita, Hisanobu},
doi = {10.2116/analsci.21.309}
}

@Article{Motz2023,
author ="Motz, Damian and Praetz, Sebastian and Schlesiger, Christopher and Henniges, Jonathan and Böttcher, Florian and Hesse, Bernhard and Castillo-Michel, Hiram and Mijatz, Steven and Malzer, Wolfgang and Kanngießer, Birgit and Vogt, Carla",
title  ="Examining iron complexes with organic ligands by laboratory XAFS",
journal  ="J. Anal. At. Spectrom.",
year  ="2023",
volume  ="38",
issue  ="2",
pages  ="391-402",
publisher  ="The Royal Society of Chemistry",
doi  ="10.1039/D2JA00351A",
url  ="http://dx.doi.org/10.1039/D2JA00351A"}

@article{Dangelo2008,
author = {D’Angelo, Paola and Lapi, Andrea and Migliorati, Valentina and Arcovito, Alessandro and Benfatto, Maurizio and Roscioni, Otello Maria and Meyer-Klaucke, Wolfram and Della-Longa, Stefano},
title = {X-ray Absorption Spectroscopy of Hemes and Hemeproteins in Solution: Multiple Scattering Analysis},
journal = {Inorganic Chemistry},
volume = {47},
number = {21},
pages = {9905-9918},
year = {2008},
doi = {10.1021/ic800982a}

}

@article{Hayakawa2004,
author = {Hayakawa, Kuniko and Hatada, Keisuke and D'Angelo, Paola and Della Longa, Stefano and Natoli, Calogero R. and Benfatto, Maurizio},
title = {Full Quantitative Multiple-Scattering Analysis of X-ray Absorption Spectra: Application to Potassium Hexacyanoferrat(II) and (III) Complexes},
journal = {Journal of the American Chemical Society},
volume = {126},
number = {47},
pages = {15618-15623},
year = {2004},
doi = {10.1021/ja045561v}
    
    

}

@article{Ross2018,
author = {Ross, Matthew and Andersen, Amity and Fox, Zachary W. and Zhang, Yu and Hong, Kiryong and Lee, Jae-Hyuk and Cordones, Amy and March, Anne Marie and Doumy, Gilles and Southworth, Stephen H. and Marcus, Matthew A. and Schoenlein, Robert W. and Mukamel, Shaul and Govind, Niranjan and Khalil, Munira},
title = {Comprehensive Experimental and Computational Spectroscopic Study of Hexacyanoferrate Complexes in Water: From Infrared to X-ray Wavelengths},
journal = {The Journal of Physical Chemistry B},
volume = {122},
number = {19},
pages = {5075-5086},
year = {2018},
doi = {10.1021/acs.jpcb.7b12532}
    
    

}

@article{Glatzel2002,
author = {Glatzel, Pieter and Jacquamet, Lilian and Bergmann, Uwe and de Groot, Frank M. F. and Cramer, Stephen P.},
title = {Site-Selective EXAFS in Mixed-Valence Compounds Using High-Resolution Fluorescence Detection: A Study of Iron in Prussian Blue},
journal = {Inorganic Chemistry},
volume = {41},
number = {12},
pages = {3121-3127},
year = {2002},
doi = {10.1021/ic010709m}

}

@article{Ito,
    author = {Ito, A. and Suenaga, M. and Ôno, K.},
    title = {Mössbauer Study of Soluble Prussian Blue, Insoluble Prussian Blue, and Turnbull's Blue},
    journal = {The Journal of Chemical Physics},
    volume = {48},
    number = {8},
    pages = {3597-3599},
    year = {1968},
    month = {04},
    issn = {0021-9606},
    doi = {10.1063/1.1669656}
}

@article{chuev_multilevel_2012,
	title = {Multilevel relaxation model for describing the Mössbauer spectra of nanoparticles in a magnetic field},
	volume = {114},
	rights = {http://www.springer.com/tdm},
	issn = {1063-7761, 1090-6509},
	url = {http://link.springer.com/10.1134/S1063776112020185},
	doi = {10.1134/S1063776112020185},
	pages = {609--630},
	number = {4},
	journaltitle = {Journal of Experimental and Theoretical Physics},
	shortjournal = {J. Exp. Theor. Phys.},
	author = {Chuev, M. A.},
	date = {2012-04},
	langid = {english},
}

@article{ERICKSON19701195,
title = {Magnetic susceptibility and Mössbauer spectrum of NH4Fe3+FeII(CN)6},
journal = {Journal of Physics and Chemistry of Solids},
volume = {31},
number = {5},
pages = {1195-1197},
year = {1970},
issn = {0022-3697},
doi = {https://doi.org/10.1016/0022-3697(70)90331-8},
url = {https://www.sciencedirect.com/science/article/pii/0022369770903318},
author = {N. Erickson and N. Elliott}
}

@article{braeuninger,
  title = {Magnetic field tuning of low-energy spin dynamics in the single-atomic magnet ${\mathrm{Li}}_{2}(\phantom{\rule{4pt}{0ex}}{\mathrm{Li}}_{1\ensuremath{-}x}{\mathrm{Fe}}_{x})\mathrm{N}$},
  author = {Br\"auninger, S. A. and Jesche, A. and Kamusella, S. and Seewald, F. and Fix, M. and Sarkar, R. and Zvyagin, A. A. and Klauss, H.-H.},
  journal = {Phys. Rev. B},
  volume = {102},
  issue = {5},
  pages = {054426},
  numpages = {17},
  year = {2020},
  month = {Aug},
  publisher = {American Physical Society},
  doi = {10.1103/PhysRevB.102.054426},
  url = {https://link.aps.org/doi/10.1103/PhysRevB.102.054426}
}

@Article{DT9760001483,
author ="Mayoh, Bryan and Day, Peter",
title  ="Charge transfer in mixed-valence solids. Part VIII. Contribution of valence delocalisation to the ferromagnetism of Prussian Blue",
journal  ="J. Chem. Soc.{,} Dalton Trans.",
year  ="1976",
issue  ="15",
pages  ="1483-1486",
publisher  ="The Royal Society of Chemistry",
doi  ="10.1039/DT9760001483",
url  ="http://dx.doi.org/10.1039/DT9760001483"}

@article{greaves_site_2016,
	title = {Site analysis and calculation of the quadrupole splitting of Prussian Blue Mössbauer spectra},
	volume = {237},
	issn = {0304-3843, 1572-9540},
	url = {http://link.springer.com/10.1007/s10751-016-1216-6},
	doi = {10.1007/s10751-016-1216-6},
	pages = {70},
	number = {1},
	journaltitle = {Hyperfine Interactions},
	shortjournal = {Hyperfine Interact},
	author = {Greaves, T. L. and Cashion, J. D.},
	urldate = {2024-11-27},
	date = {2016-12},
	langid = {english},
}

@article{kamusella,
	title = {Moessfit: {A} free {Mössbauer} fitting program},
	volume = {237},
	copyright = {http://www.springer.com/tdm},
	issn = {0304-3843, 1572-9540},
	shorttitle = {Moessfit},
	url = {http://link.springer.com/10.1007/s10751-016-1247-z},
	doi = {10.1007/s10751-016-1247-z},
	language = {en},
	number = {1},
	urldate = {2025-07-14},
	journal = {Hyperfine Interactions},
	author = {Kamusella, Sirko and Klauss, Hans-Henning},
	month = dec,
	year = {2016},
	note = {Publisher: Springer Science and Business Media LLC},
}

@Article{Schlesiger2020,
author ="Schlesiger, Christopher and Praetz, Sebastian and Gnewkow, Richard and Malzer, Wolfgang and Kanngießer, Birgit",
title  ="Recent progress in the performance of HAPG based laboratory EXAFS and XANES spectrometers",
journal  ="J. Anal. At. Spectrom.",
year  ="2020",
volume  ="35",
issue  ="10",
pages  ="2298-2304",
publisher  ="The Royal Society of Chemistry",
doi  ="10.1039/D0JA00208A",
url  ="http://dx.doi.org/10.1039/D0JA00208A"}

@Article{Schlesiger2015,
author ="Schlesiger, C. and Anklamm, L. and Stiel, H. and Malzer, W. and Kanngießer, B.",
title  ="XAFS spectroscopy by an X-ray tube based spectrometer using a novel type of HOPG mosaic crystal and optimized image processing",
journal  ="J. Anal. At. Spectrom.",
year  ="2015",
volume  ="30",
issue  ="5",
pages  ="1080-1085",
publisher  ="The Royal Society of Chemistry",
doi  ="10.1039/C4JA00303A",
url  ="http://dx.doi.org/10.1039/C4JA00303A"}

@article{Ravel2005,
author = {Ravel, B. and Newville, M.},
title = {ATHENA, ARTEMIS, HEPHAESTUS: data analysis for X-ray absorption spectroscopy using IFEFFIT},
journal = {Journal of Synchrotron Radiation},
volume = {12},
number = {4},
pages = {537-541},
doi = {https://doi.org/10.1107/S0909049505012719},
year = {2005}
}

@article{Praetz2025,
    doi = {10.1371/journal.pone.0323678},
    author = {Praetz, Sebastian AND Schlesiger, Christopher AND Motz, Damian Alexander AND Klimke, Stephen AND Jahns, Moritz AND Gottschalk, Christine AND Heinrich, Lena AND Heppke, Eva Maria AND Malzer, Wolfgang AND Renz, Franz AND Vogt, Carla AND Kanngießer, Birgit},
    journal = {PLOS ONE},
    publisher = {Public Library of Science},
    title = {Can laboratory-based XAFS compete with XRD and Mössbauer spectroscopy as a tool for quantitative species analysis? Critical evaluation using the example of a natural iron ore},
    year = {2025},
    month = {05},
    volume = {20},
    pages = {1-29},
    number = {5},

}

@Article{Gili2024,
author ="Gili, Albert and Bekheet, Maged F. and Thimm, Franziska and Bischoff, Benjamin and Geske, Michael and Konrad, Martin and Praetz, Sebastian and Schlesiger, Christopher and Selve, Sören and Gurlo, Aleksander and Rosowski, Frank and Schomäcker, Reinhard",
title  ="One-pot synthesis of iron-doped ceria catalysts for tandem carbon dioxide hydrogenation",
journal  ="Catal. Sci. Technol.",
year  ="2024",
volume  ="14",
issue  ="15",
pages  ="4174-4186",
publisher  ="The Royal Society of Chemistry",
doi  ="10.1039/D4CY00439F",
url  ="http://dx.doi.org/10.1039/D4CY00439F"}

@article{Oliveira2023,
author = {Lima Oliveira, Rafael and Ledwa, Karolina A. and Chernyayeva, Olga and Praetz, Sebastian and Schlesiger, Christopher and Kepinski, Leszek},
title = {Cerium Oxide Nanoparticles Confined in Doped Mesoporous Carbons: A Strategy to Produce Catalysts for Imine Synthesis},
journal = {Inorganic Chemistry},
volume = {62},
number = {33},
pages = {13554-13565},
year = {2023},
doi = {10.1021/acs.inorgchem.3c01985}
    
    

}

@book{Jander2006,
  author		= "Strähle, J. and Schweda, E.",
  title			= "Jander/Blasius - Lehrbuch der analytischen und präparativen anorganischen Chemie",
  address		= "Stuttgart",
  publisher		= "S. Hirzel Verlag",
  year			= "2006"
}

@article{hoveeffect1990,
	title = {Effect of {Prussian} blue (ammonium-iron-hexacyanoferrate) in reducing the accumulation of radiocesium in reindeer},
	volume = {10},
	copyright = {http://creativecommons.org/licenses/by/3.0/},
	issn = {1890-6729},
	doi = {10.7557/2.10.3.820},
	number = {3},
	journal = {Rangifer},
	author = {Hove, K. and Staaland, H. and Pedersen, K. and Sletten, H. D.},
	month = sep,
	year = {1990},
	note = {Publisher: UiT The Arctic University of Norway},
	pages = {43},
}

@article{mathiesen1990,
	title = {Elimination of radiocesium in contaminated adult female {Norwegian} reindeer},
	volume = {10},
	copyright = {http://creativecommons.org/licenses/by/3.0/},
	issn = {1890-6729},
	doi = {10.7557/2.10.3.823},
	number = {3},
	journal = {Rangifer},
	author = {Mathiesen, S. D. and Nordøy, L. M. and Blix, A. S.},
	month = sep,
	year = {1990},
	note = {Publisher: UiT The Arctic University of Norway},
	pages = {49},
}

@article{Dresow,
author = {Bernd Dresow and Peter Nielsen and Roland Fischer and Alexander A. Pfau and Hellmuth H. Heinrich},
title = {In vivo Binding of Radiocesium by Two Forms of Prussian Blue and by Ammonium Iron Hexacyanoferrate (ii)},
journal = {Journal of Toxicology: Clinical Toxicology},
volume = {31},
number = {4},
pages = {563--569},
year = {1993},
publisher = {Taylor \& Francis},
doi = {10.3109/15563659309025761},
    
    

}

@inbook{Leng2013,
author = {Leng, Yang},
publisher = {John Wiley and Sons, Ltd},
isbn = {9783527670772},
title = {Light Microscopy},
booktitle = {Materials Characterization},
chapter = {1},
pages = {1-45},
doi = {https://doi.org/10.1002/9783527670772.ch1},
year = {2013},
address = {New York}
}

@Inbook{Rochow1978,
author="Rochow, Theodore George
and Rochow, Eugene George",
title="Compound Microscopes Using Reflected Light",
bookTitle="An Introduction to Microscopy by Means of Light, Electrons, X-Rays, or Ultrasound",
year="1978",
publisher="Springer US",
address="Boston, MA",
pages="63--90",
isbn="978-1-4684-2454-6",
doi="10.1007/978-1-4684-2454-6_4"
}

@book{Borchardt-Ott2009,
title="Kristallographie: Eine Einf{\"u}hrung f{\"u}r Naturwissenschaftler",
author= "Borchardt-Ott, Walter",
year="2009",
publisher="Springer Berlin Heidelberg",
address="Berlin, Heidelberg",
pages="281--297",
isbn="978-3-540-78271-1",
doi="10.1007/978-3-540-78271-1_13",
chapter = "13"
}

@article{Sharma2012, 
title={X-ray diffraction: a powerful method of characterizing nanomaterials}, volume={4}, url={https://updatepublishing.com/journal/index.php/rrst/article/view/933}, abstractNote={&amp;lt;p class=&amp;quot;MsoNormal&amp;quot; style=&amp;quot;text-align: justify;&amp;quot;&amp;gt;X-ray diffraction techniques are a very useful characterization tool to study, non-destructively, the crystallographic structure, chemical composition and physical properties of materials and thin films. It can also be used to measure various structural properties of these crystalline phases such as strain, grain size, phase composition, and defect structure. XRD is also used to determine the thickness of thin films, as well as the atomic arrangements in amorphous materials such as polymers. This paper reports the importance of X-ray diffraction technique for the characterization of nanomaterials.&amp;lt;/p&amp;gt;}, number={8}, journal={Recent Research in Science and Technology}, author={Sharma, Ravi and Bisen, D P and Shukla, Usha and Sharma, B G}, year={2012}, month={Oct.} 
}

@Article{Ameh2019,
author={Ameh, E. S.},
title={A review of basic crystallography and x-ray diffraction applications},
journal={The International Journal of Advanced Manufacturing Technology},
year={2019},
month={Dec},
day={01},
volume={105},
number={7},
pages={3289-3302},
issn={1433-3015},
doi={10.1007/s00170-019-04508-1},
url={https://doi.org/10.1007/s00170-019-04508-1}
}

@phdthesis{Motz2021,
    author ={Motz, Damian Alexander} ,
    title = {Entwicklung von Referenzmaterialien für die Röntgen-Nahkanten-Absorptionsspektroskopie am Laboraufbau},
    school = {Leibniz Universität Hannover},
    year = 2021,
    doi = {https://doi.org/10.15488/10431}
}

@article{gieserev,
title = {Countermeasures for reducing the transfer of radiocesium to animal derived foods},
journal = {Science of The Total Environment},
volume = {85},
pages = {317-327},
year = {1989},
note = {Transfer of Radionuclides to Livestock},
issn = {0048-9697},
doi = {https://doi.org/10.1016/0048-9697(89)90331-8},
url = {https://www.sciencedirect.com/science/article/pii/0048969789903318},
author = {W.W. Giese}
}

@article{koehler,
author = {K{\"o}hler, Frank H. and Storcheva, Oksana},
title = {Paramagnetic Prussian Blue Analogues CsMII[MIII(CN)6]. The Quest for Spin on Cesium Ions by Use of 133Cs MAS NMR Spectroscopy},
journal = {Inorganic Chemistry},
volume = {54},
number = {14},
pages = {6801-6806},
year = {2015},
doi = {10.1021/acs.inorgchem.5b00711}
}

@Article{Praetz2025_2,
author ="Praetz, Sebastian and Johansen, Morten and Kober, Delf and Tesic, Marko and Schlesiger, Christopher and Bomholdt Ravnsbæk, Dorthe and Kanngießer, Birgit",
title  ="Operando laboratory XAS on battery materials using the DANOISE cell in a von Hámos spectrometer",
journal  ="J. Anal. At. Spectrom.",
year  ="2025",
pages  ="-",
publisher  ="The Royal Society of Chemistry",
doi  ="10.1039/D5JA00155B"}

@article{Praetz2025_3,
    author = {Praetz, S. and Grötzsch, D. and Schlesiger, C. and Motz, D. and Würth, M. and Zimmermann, R. and Lucka, R. and Malzer, W. and Lützenkirchen-Hecht, D. and Renz, F. and Kanngießer, B.},
    title = {In situ heating cell for temperature dependent transmission x-ray absorption spectroscopy (XAS) measurement with a laboratory based spectrometer},
    journal = {Review of Scientific Instruments},
    volume = {96},
    number = {3},
    pages = {035120},
    year = {2025},
    month = {03},
    issn = {0034-6748},
    doi = {10.1063/5.0253653}
}

@article{quin,
author = {Qin, Zhiguo and Li, Yan and Gu, Ning},
title = {Progress in Applications of Prussian Blue Nanoparticles in Biomedicine},
journal = {Advanced Healthcare Materials},
volume = {7},
number = {20},
pages = {1800347},
doi = {https://doi.org/10.1002/adhm.201800347},
year = {2018}
}

@article{zhang_2025,
	title = {Multidimensional applications of prussian blue-based nanoparticles in cancer immunotherapy},
	volume = {23},
	issn = {1477-3155},
	url = {https://jnanobiotechnology.biomedcentral.com/articles/10.1186/s12951-025-03236-x},
	doi = {10.1186/s12951-025-03236-x},
	language = {en},
	number = {1},
	journal = {Journal of Nanobiotechnology},
	author = {Zhang, Jiayi and Wang, Fang and Sun, Zhaogang and Ye, Jun and Chu, Hongqian},
	month = mar,
	year = {2025},
	pages = {161}
}

@Article{gao,
AUTHOR = {Gao, Xiaoran and Wang, Qiaowen and Cheng, Cui and Lin, Shujin and Lin, Ting and Liu, Chun and Han, Xiao},
TITLE = {The Application of Prussian Blue Nanoparticles in Tumor Diagnosis and Treatment},
JOURNAL = {Sensors},
VOLUME = {20},
YEAR = {2020},
NUMBER = {23},
ARTICLE-NUMBER = {6905},
URL = {https://www.mdpi.com/1424-8220/20/23/6905},
PubMedID = {33287186},
ISSN = {1424-8220},
DOI = {10.3390/s20236905}
}

@article{busquet,
title = {Prussian blue nanoparticles: synthesis, surface modification, and biomedical applications},
journal = {Drug Discovery Today},
volume = {25},
number = {8},
pages = {1431-1443},
year = {2020},
issn = {1359-6446},
doi = {https://doi.org/10.1016/j.drudis.2020.05.014},
url = {https://www.sciencedirect.com/science/article/pii/S1359644620302063},
author = {Maria Antònia Busquets and Joan Estelrich}
}

@misc{DIN-9276-1,
    author = {DIN},
    type = {Norm},
    number = {DIN ISO 9276-1},
    year = {2004},
    month = {09},
    title = {Darstellung der Ergebnisse von Partikelgrößenanalysen - Teil 1: Grafische Darstellung (ISO 9276-1:1998) [Representation of results of particle size analysis - Part 1: Graphical representation (ISO 9276-1:1998)]},
    doi = {https://dx.doi.org/10.31030/9560648}
}

@misc{DIN-9276-2,
    author = {DIN},
    type = {Norm},
    number = {DIN ISO 9276-2},
    year = {2018},
    month = {09},
    title = {Darstellung der Ergebnisse von Partikelgrößenanalysen - Teil 2: Berechnung von mittleren Partikelgrößen/-durchmesern und Momenten aus Partikelgrößenverteilungen (ISO 9276-2:2014) [Representation of results of particle size analysis - Part 2: Calculation of average particle sizes/diameters and moments from particle size distributions (ISO 9276-2:2014)]},
    doi = {https://dx.doi.org/10.31030/2867011}
}

@misc{DIN-13320,
    author = {DIN},
    type = {Norm},
    number = {DIN ISO 13320},
    year = {2022},
    month = {12},
    title = {Partikelgrößenanalyse - Laserbeugungsverfahren (ISO 13320:2020) [Particle size analysis - Laser diffraction methods (ISO 13320:2020)]},
    doi = {https://dx.doi.org/10.31030/3384453}
}

@misc{DIN-38402,
    author = {DIN},
    type = {Norm},
    number = {DIN 38402-51},
    year = {2017},
    month = {05},
    title = {Deutsche Einheitsverfahren zur Wasser-, Abwasser- und Schlammuntersuchung - Allgemeine Angaben (Gruppe A) - Teil 51: Kalibrierung von Analysenverfahren - Lineare Kalibrierfunktion (A 51) German standard methods for the examination of water, waste water and sludge - General information (group A) - Part 51: Calibration of analytical methods - Linear calibration (A 51)},
    doi = {https://dx.doi.org/10.31030/2657448}
}

@article{middlemiss,
  title = {Ferromagnetism and spin transitions in prussian blue: A solid-state hybrid functional study},
  author = {Middlemiss, Derek S. and Wilson, Chick C.},
  journal = {Phys. Rev. B},
  volume = {77},
  issue = {15},
  pages = {155129},
  numpages = {13},
  year = {2008},
  month = {Apr},
  publisher = {American Physical Society},
  doi = {10.1103/PhysRevB.77.155129},
  url = {https://link.aps.org/doi/10.1103/PhysRevB.77.155129}
}

@article{SONODA2025,
title = {Perls' Prussian blue staining and chemistry of Prussian blue and Turnbull blue},
journal = {Forensic Science International: Synergy},
volume = {11},
pages = {100627},
year = {2025},
issn = {2589-871X},
doi = {https://doi.org/10.1016/j.fsisyn.2025.100627},
url = {https://www.sciencedirect.com/science/article/pii/S2589871X25000567},
author = {Ai Sonoda and Masayuki Nihei and Norihiro Shinkawa and Eiji Kakizaki and Nobuhiro Yukawa}
}

@book{Lurie1978,
     author = {Ju. Lurie},
     title = {Handbook of Analytical Chemistry},
     publisher = {Mir Publishers},
     year = {1978},
    pages = {105-116},
   address = {Moscow}
 }

@article{Dostal1996,
title = {Lattice contractions and expansions accompanying the electrochemical conversions of Prussian blue and the reversible and irreversible insertion of rubidium and thallium ions},
journal = {Journal of Electroanalytical Chemistry},
volume = {406},
number = {1},
pages = {155-163},
year = {1996},
issn = {1572-6657},
doi = {https://doi.org/10.1016/0022-0728(95)04427-2},
url = {https://www.sciencedirect.com/science/article/pii/0022072895044272},
author = {Aleš Dostal and Günther Kauschka and S.Jayarama Reddy and Fritz Scholz}
}

@article{MAZEI2011,
title = {Electrocatalytic reduction of hydrogen peroxide at Prussian blue modified electrode: An in situ Raman spectroelectrochemical study},
journal = {Journal of Electroanalytical Chemistry},
volume = {660},
number = {1},
pages = {140-146},
year = {2011},
issn = {1572-6657},
doi = {https://doi.org/10.1016/j.jelechem.2011.06.022},
author = {Regina Mažeikienė and Gediminas Niaura and Albertas Malinauskas}
}

@article{barsam2011,
author = {Barsan, Mirela M. and Butler, Ian S. and Fitzpatrick, Jessica and Gilson, Denis F. R.},
title = {High-pressure studies of the micro-Raman spectra of iron cyanide complexes: Prussian blue (Fe4[Fe(CN)6]3), potassium ferricyanide (K3[Fe(CN)6]), and sodium nitroprusside (Na2[Fe(CN)5(NO)]·2H2O)},
journal = {Journal of Raman Spectroscopy},
volume = {42},
number = {9},
pages = {1820-1824},
doi = {https://doi.org/10.1002/jrs.2931},
year = {2011}
}

\end{document}